\documentclass[12pt]{article}

\usepackage{graphicx}
\usepackage{amsmath, amssymb}

\begin{document}
  
\begin{titlepage}

\def\thefootnote{\fnsymbol{footnote}}

\begin{center}

\hfill TU-818 \\
\hfill UT-HET 009 \\
\hfill May, 2008

\vspace{0.5cm}
{\Large\bf  High Energy Cosmic Rays \\
from the Decay of Gravitino Dark Matter}

\vspace{1cm}
{\large Koji Ishiwata}$^{\it (a)}$\footnote{E-mail: 
ishiwata@tuhep.phys.tohoku.ac.jp},
{\large Shigeki Matsumoto}$^{\it (b)}$\footnote{E-mail: 
smatsu@sci.u-toyama.ac.jp},
{\large Takeo Moroi}$^{\it (a)}$\footnote{E-mail: 
moroi@tuhep.phys.tohoku.ac.jp}

\vspace{1cm}

{\it $^{(a)}${Department of Physics, Tohoku University,
    Sendai 980-8578, Japan}}

\vspace{0.5cm}

{\it $^{(b)}${Department of Physics, University of Toyama, 
    Toyama 930-8555, Japan}}

\vspace{1cm}
\abstract{ 

  We study high energy cosmic rays from the decay of the gravitino
  dark matter in the framework of supersymmetric model with $R$-parity
  violation.  Even though $R$-parity is violated, the lifetime of the
  gravitino, which is assumed to be the lightest superparticle, can be
  longer than the present age of the universe if $R$-parity violating
  interactions are weak enough.  We have performed a detailed
  calculation of the fluxes of gamma ray and positron from the decay
  of the gravitino dark matter.  We also discuss the implication of
  such a scenario to present and future observations of high energy
  cosmic rays.  In particular, we show that the excess of the
  gamma-ray flux observed by EGRET and the large positron fraction
  observed by HEAT can be simultaneously explained by the cosmic rays
  from the decay of the gravitino dark matter.

 }

\end{center}
\end{titlepage}

\renewcommand{\theequation}{\thesection.\arabic{equation}}
\renewcommand{\thepage}{\arabic{page}}
\setcounter{page}{1}
\renewcommand{\thefootnote}{\#\arabic{footnote}}
\setcounter{footnote}{0}

\section{Introduction}
\label{sec:intro}
\setcounter{equation}{0}

In particle cosmology, origin of dark matter of the universe is one of
the most important problems.  Since there is no viable candidate for
dark matter in the particle content of the standard model, new physics
beyond the standard model is necessary to solve this problem.
Supersymmetric model is a prominent candidate for the physics beyond
the standard model; it not only introduces a viable candidate for dark
matter, which is the lightest superparticle (LSP), but also solves
other serious problems in particle physics, like naturalness problem
of the electro-weak symmetry breaking.

In order to realize LSP dark matter, conservation of $R$-parity is
usually assumed.  In \cite{Buchmuller:2007ui}, however, it was pointed
out that LSP dark matter scenario may be realized even with $R$-parity
violation (RPV) if the LSP is the gravitino; even though gravitino
LSP becomes unstable with RPV, its lifetime may be longer than the
present age of the universe because the decay rate of the gravitino is
suppressed by the Planck mass as well as by (small) RPV parameter.
Such a scenario has a great advantage for the thermal leptogenesis
scenario \cite{Fukugita:1986hr}, as we will briefly discuss in the
next section.  Then, the primordial gravitino produced in the early
universe can be a viable candidate for dark matter.

Even though the lifetime of the gravitino is much longer than the
present age of the universe, a fraction of the gravitinos have decayed
until today.  Such a decay becomes a source of high energy cosmic rays
\cite{Buchmuller:2007ui,Ibarra:2007wg}; the decay of the gravitino
dark matter may produce high energy gamma ray and positron, which may
be observed by present and future experiments.

In this paper, we investigate gamma-ray and positron fluxes from the
decay of gravitino dark matter in supersymmetric model with RPV.  For
this purpose, we first calculate the decay rate and branching ratios
of the gravitino taking into account all the relevant operators.
Fragmentation and hadronization of the decay products are studied by
using PYTHIA package \cite{pythia}.  Then, we calculate fluxes of
gamma ray and positron from the decay of the gravitino dark matter,
carefully taking account of the propagation of the cosmic rays.  We
discuss the implications of our results to the present and future
observations of the high energy cosmic rays.  Importantly, excesses of
the gamma-ray and positron fluxes over the backgrounds are reported by
Energetic Gamma Ray Experiment Telescope (EGRET) \cite{egret} and High
Energy Antimatter Telescope (HEAT) \cite{heat} experiments,
respectively.  We will show that these excesses may be simultaneously
explained by the scenario mentioned above.

There are many past works to address HEAT and EGRET anomalies in terms
of the annihilation of weakly-interacting-massive-particle (WIMP) dark
matter
\cite{Jungman:1995df,Baltz:1998xv,AnnihilationPastWorks,Hooper:2004bq,Hisano:2005ec}.
However, the predictions of the present scenario significantly differ
from those of the annihilation scenarios.  First, it is difficult to
explain these anomalies simultaneously in the annihilation scenario; it
is quite unlikely that dark matter annihilation is a main constituent of
extragalactic gamma ray without exceeding the observed gamma-ray flux
from the Galactic center \cite{Ando:2005hr}. On the other hand, in the
case of the decaying dark matter, the above constraint is relaxed,
because the production rate of the gamma-ray is proportional not to the
square of the number density of the dark matter but to the
density. Then, as we will show, it is also possible to explain HEAT and
EGRET anomalies simultaneously.  It is also notable that, in the
decaying dark matter scenario, the fluxes of the high energy cosmic rays
are insentive to the boost factor, on which the fluxes in the
annihilation scenarios strongly depend on.

The organization of this paper is as follows.  In Section
\ref{sec:cosmo}, we first summarize the cosmological scenario that we
consider.  In Section \ref{sec:decay}, we discuss the decay processes
of the gravitino.  In Section \ref{sec:formalism}, formulae to
calculate the cosmic-ray fluxes are given.  The gamma-ray and positron
fluxes from the decay of the gravitino dark matter are shown in
Section \ref{sec:results}; readers who are mainly interested in the
results may directly go to this section.  Section \ref{sec:discussion}
is devoted to conclusions and discussion.

\section{Cosmological Scenario}
\label{sec:cosmo}
\setcounter{equation}{0}

We first introduce the cosmological scenario that we consider.
Although $R$-parity conservation is usually assumed in conventional
studies of supersymmetric models, RPV has a favourable aspect in
cosmology.  In supersymmetric models, it is often the case that the
thermal leptogenesis scenario \cite{Fukugita:1986hr}, which is one of
the most prominent scenario to generate the present baryon asymmetry
of the universe, is hardly realized since such a scenario requires
relatively high reheating temperature after inflation, $T_{\rm
  R}\gtrsim 10^9\ {\rm GeV}$ \cite{Buchmuller:2004nz,Giudice:2003jh}.
With such a high reheating temperature, gravitino overproduction
problem arises for wide range of the gravitino mass as far as
$R$-parity is conserved \cite{Kawasaki:2008qe}.  If the gravitino is
unstable, gravitino produced after the reheating decays after the
big-bang nulceosynthesis (BBN) starts and spoils the success of the
BBN.\footnote
{However, $T_{\rm R}\sim 10^9\ {\rm GeV}$ may be also allowed when the
gravitino mass is larger than $O(10\ {\rm TeV})$.  Such a scenario may
be realized in the class of anomaly-mediation model
\cite{Giudice:1998xp,Randall:1998uk}.}
If the gravitino is stable, on the contrary, the primordial gravitino
survives until today and contributes to the present energy density of
the universe.  Since the gravitino abundance increases as the
gravitino mass becomes smaller, overclosure of the universe happens
unless the gravitino mass is large enough \cite{Moroi:1993mb}.  With
$T_{\rm R}\gtrsim 10^9\ {\rm GeV}$, the above problems can be avoided
only when (i) the gravitino is stable, and (ii) the gravitino mass is
around $100\ {\rm GeV}$.  However, even in such a case, one has to
worry about the decay of the lightest superparticle in the minimal
supersymmetric standard model (MSSM) sector, which we call MSSM-LSP;
with $R$-parity conservation, the MSSM-LSP, which is assumed to be the
next-to-the-lightest superparticle (NLSP), decays only into gravitino
and some standard-model particle(s).  When $m_{3/2}\sim 100\ {\rm
  GeV}$, the lifetime of the MSSM-LSP becomes longer than $1\ {\rm
  sec}$ and relic MSSM-LSP decays after the BBN starts.  When the
MSSM-LSP is the neutralino or charged slepton, such decay processes
spoil the success of the BBN, and hence it is difficult to realize the
thermal leptogenesis scenario.

If the $R$-parity is violated, the MSSM-LSP may decay via RPV
interaction and its lifetime may become shorter than $1\ {\rm sec}$.
Then, $T_{\rm R}\sim 10^9\ {\rm GeV}$ is allowed.  In such a case,
gravitino is no longer stable and decays to standard-model particles.
Even in such a case, however, the lifetime of the gravitino may be
longer than the present age of the universe and hence the gravitino
dark matter and thermal leptogenesis may be simultaneously realized
\cite{Buchmuller:2007ui}.

Here, we consider the case where the gravitino is the LSP in the
framework of the $R$-parity violated supersymmetric models.  We assume
that the present mass density of the gravitino is equal to the
observed dark matter density so that the gravitino can play the role
of dark matter.  There are several possibilities of the origin of such
a primordial gravitino: scattering processes of thermal particles
\cite{Moroi:1993mb} or the decay of scalar condensations
\cite{GravFromScalars}.  We will not discuss in detail about the
production mechanism of the primordial gravitino because the following
arguments hold irrespective of the origin of the gravitino.  In
addition, we consider the case that the lifetime of the gravitino is
much longer than the present age of the universe; the upper bounds on
the size of RPV couplings will be discussed in the following
section.

\section{Framework and Decay Rates}
\label{sec:decay}
\setcounter{equation}{0}

In this section, we introduce the supersymmetric model that we
consider.  Then, we summarize the decay rates of the gravitino and the
NLSP, which are important for our study.

\subsection{Model}

In this article, we concentrate on the case where the $R$-parity
violating interactions originate from bi-linear terms of Higgs and
lepton doublet.  In the original basis, the $R$-parity violating
interactions are assumed to be bi-linear terms in superpotential and
supersymmetry (SUSY) breaking terms \cite{Roy:1996bua}.  Without loss
of generality, we can always eliminate the bi-linear $R$-parity
violating terms from the superpotential by the redefinition of the
Higgs and lepton-doublet multiplets.  Then, the mixing terms between
the Higgsino and lepton doublets are eliminated from the fermion mass
matrix.  In the following, we work in such a basis.  Then, the
relevant $R$-parity violating terms are only in the soft-SUSY breaking
terms, which are given by
\begin{eqnarray}
  {\cal L}_{\rm RPV} 
  = B_i \tilde{L}_i H_u + m^2_{\tilde{L}_i H_d} \tilde{L}_i H^*_d 
  + {\rm h.c.},
  \label{L_RPV}
\end{eqnarray}
where $\tilde{L}_i$ is left-handed slepton doublet in $i$-th
generation, while $H_u$ and $H_d$ are up- and down-type Higgs boson
doublets, respectively.  In the following, we study the
phenomenological consequences of the $R$-parity violating terms given
in Eq.\ (\ref{L_RPV}).

With these $R$-parity violating terms, the vacuum expectation values
(VEVs) of left-handed sneutrino fields $\tilde{\nu}_{i}$ are
generated as
\begin{eqnarray}
  \langle \tilde{\nu}_{i} \rangle
  = \frac{B_i \sin\beta + m^2_{\tilde{L}_i H_d} \cos\beta}
  {m^2_{\tilde{\nu}_{i}}} v,
\end{eqnarray}
where $v\simeq 174\ {\rm GeV}$ is the VEV of standard-model-like Higgs
boson, $\tan \beta = \langle H^0_u \rangle / \langle H^0_d \rangle$,
and $m_{\tilde{\nu}_{i}}$ is the mass of $\tilde{\nu}_{i}$.  The VEVs
of the sneutrinos play important role in the following analysis.  We
parametrize the VEVs of the sneutrinos as
\begin{eqnarray}
  \kappa_i \equiv \frac{\langle \tilde{\nu}_i \rangle}{v},
\end{eqnarray}
and consider the case that $\kappa_i\ll 1$.

One important constraint on the size of the $R$-parity violation is
from the neutrino masses.  Here, we assume that the neutrino masses
are mainly from some other interaction, like the seesaw mechanism
\cite{seesaw} or Dirac-type Yukawa interaction.  However, the VEVs of
sneutrinos also generate neutrino masses; assuming the Majorana-type
masses for neutrinos, the $ij$ component of the mass matrix receives
the contribution of
\begin{eqnarray}
  \left[ \Delta m_{\nu} \right]_{ij} = 
  m_Z^2 
  \kappa_i \kappa_j
  \sum_{\alpha} \frac{|c_{\tilde{Z} \tilde{\chi}^0_{\alpha}}|^2}
  {m_{\tilde{\chi}^0_{\alpha}}},
  \label{eq:numass}
\end{eqnarray}
where $m_Z$ is the $Z$-boson mass.  In addition, Zino $\tilde{Z}$,
which is the superpartner of the $Z$-boson, is related to the mass
eigenstates of the neutralinos $\tilde{\chi}^0_{\alpha}$ (with mass
$m_{\tilde{\chi}^0_{\alpha}}$) as
\begin{eqnarray}
  \tilde{Z} = \sum_\alpha c_{\tilde{Z}\tilde{\chi}^0_{\alpha}} 
  \tilde{\chi}^0_{\alpha}.
\end{eqnarray}
(We also define the coefficients for photino and Higgsinos,
$c_{\tilde{\gamma}\tilde{\chi}^0_{\alpha}}$,
$c_{\tilde{H}_u^0\tilde{\chi}^0_{\alpha}}$, and
$c_{\tilde{H}_d^0\tilde{\chi}^0_{\alpha}}$, by replacing
$\tilde{Z}\rightarrow\tilde{\gamma}$, $\tilde{H}_u^0$, and
$\tilde{H}_d^0$.)  Assuming that the neutralino masses are close to
the electro-weak scale so that the SUSY can be the solution to the
fine-tuning problem of the electro-weak symmetry breaking, the
correction to the neutrino mass matrix is estimated as
\begin{eqnarray}
  \left[ \Delta m_{\nu} \right]_{ij}
  \sim 10^{-3} \ {\rm eV} \times
  \left( \frac{\kappa_i}{10^{-7}} \right)
  \left( \frac{\kappa_j}{10^{-7}} \right).
\end{eqnarray}
It indicates that the $R$-parity induced neutrino mass does not exceed
experimental bound of observed neutrino mass when
$\kappa_i\lesssim 10^{-7}$ is satisfied.

As we have mentioned, one of the important motivations to consider
RPV is to relax the BBN constraints due to the decay of the MSSM-LSP.
In a case that MSSM-LSP is Bino-like neutralino $\tilde{B}$, it decays
in two-body processes, $\tilde{B} \rightarrow Z \nu_i$, $Wl_i$, and
$h\nu_i$.  The decay rates of each mode are given by\footnote
{Here, we consider the case that the lightest Higgs boson $h$ is almost
standard-model like, so that the Higgs mixing angle is given by the
$\beta$ parameter.}
\begin{eqnarray}
  \Gamma_{\tilde{B} \rightarrow Z \nu_i}
  &=&
  \frac{1}{128 \pi}g^2_Z \sin^2 \theta_W \kappa^2_i \ 
  m_{\tilde{B}}
  \left(
  1-3\frac{m^4_Z}{m^4_{\tilde{B}}}+2\frac{m^6_Z}{m^6_{\tilde{B}}}
  \right),
\\
  \Gamma_{\tilde{B} \rightarrow W l_i}
  &=&
  \frac{1}{64 \pi} g^2_Z \sin^2 \theta_W \kappa^2_i \ 
  m_{\tilde{B}}
  \left(
  1-3\frac{m^4_W}{m^4_{\tilde{B}}}+2\frac{m^6_W}{m^6_{\tilde{B}}}
  \right),
\\
  \Gamma_{\tilde{B} \rightarrow h \nu_i}
  &=&
  \frac{1}{128 \pi} g^2_Z \sin^2 \theta_W \kappa^2_i \ 
  m_{\tilde{B}}
  \left(\frac{m^2_{\tilde{\nu}}}{m^2_{\tilde{\nu}}-m^2_h}
  \right)^2
  \left(
  1-\frac{m^2_h}{m^2_{\tilde{B}}}
  \right)^2,
\end{eqnarray}
where $g_Z=\sqrt{g_1^2+g_2^2}$ (with $g_1$ and $g_2$ being the gauge
coupling constants of the $U(1)_Y$ and $SU(2)_L$ gauge groups,
respectively), $\theta_W$ is the Weinberg angle, $m_{\tilde{B}}$ is
Bino-like neutralino mass, and $m_X$ ($X=Z$, $W$, $h$) is the mass of
gauge or Higgs boson.  Hence the lifetime is estimated as
\begin{eqnarray}
  \tau_{\tilde{B}} \simeq
  0.01\ {\rm sec} \times
  \left( \frac{\kappa}{10^{-11}} \right)^{-2} 
  \left( \frac{m_{\tilde{B}}}{200~{\rm GeV}} \right)^{-1},
\end{eqnarray}
where
\begin{eqnarray}
  {\kappa}^2 \equiv \sum_i \kappa_i^2.
\end{eqnarray}
In another case that MSSM-LSP is right-handed stau $\tilde{\tau}_R$,
it decays in the processes, $\tilde{\tau}_R\rightarrow\tau\nu_i$.  In
such a case, the decay rate is given by
\begin{eqnarray}
  \Gamma_{\tilde{\tau}_R}
  = 
  \frac{1}{16 \pi}g^4_Z \sin^4 \theta_W \kappa^2
  \left(
  \frac{v}{m_{\tilde{\chi}^0}}
  \right)^2
  m_{\tilde{\tau}_R},
\end{eqnarray}
where $m_{\tilde{\chi}^0}$ and $m_{\tilde{\tau}_R}$ masses of the
lightest neutralino and stau, respectively, and the lifetime is
estimated as
\begin{eqnarray}
  \tau_{\tilde{\tau}_R} \simeq
  0.3\ {\rm sec} \times
  \left( \frac{\kappa}{10^{-11}} \right)^{-2}
  \left( \frac{m_{\tilde{\chi}^0}}{300~{\rm GeV}} \right)^2
  \left( \frac{m_{\tilde{\tau}_R}}{200\ {\rm GeV}} \right)^{-1}.
\end{eqnarray}
Therefore, the lifetime of the MSSM-LSP, which is the NLSP in this
case, becomes shorter than $\sim 1\ {\rm sec}$ if typically $\kappa_i
\gtrsim 10^{-11}$ is satisfied.  Thus, combined with the upper bound
for $\kappa_i$ from the neutrino mass, we focus on the parameter
region:
\begin{eqnarray}
  10^{-11} \lesssim \kappa_i \lesssim 10^{-7}.
\end{eqnarray}

Before closing this subsection, we comment on the effects of
tri-linear $R$-parity violating terms induced by the redefinition of
the Higgs and lepton-doublet multiplets.  With the redefinition of
$H_d$ and $L_i$ to eliminate the bi-linear $R$-parity violating terms
from the superpotential, tri-linear $R$-parity violating terms are
induced.  They are irrelevant for our following studies, but are
constrained, in particular, from the wash-out of the baryon asymmetry
of the universe.\footnote
{In the present setup, baryon number is conserved, so the constraints
  from the nucleon decays are irrelevant.}
Let us denote the tri-linear $R$-parity violating terms in the
superpotential as
\begin{eqnarray}
  W_{\rm RPV}
  =
  \lambda_{ijk}\hat{L}_k \hat{L}_i \hat{E}^c_j + 
  \lambda'_{ijk}\hat{L}_k \hat{Q}_i \hat{D}^c_j, 
\end{eqnarray}
where $\hat{L}_i,$ and $\hat{Q}_i$ are left-handed lepton, quark
doublets, while $\hat{E}^c_i$ and $\hat{D}^c_i$ are right-handed
lepton, down-quark singlets, respectively.  (Here, ``hat'' is for
superfield.)  Then, in order not to wash out the baryon asymmetry of
the universe, the coupling constants in the above superpotential are
constrained as \cite{Campbell:1990fa}
\begin{eqnarray}
  \lambda_{ijk}, \lambda'_{ijk} \lesssim 10^{-7}.
\end{eqnarray}
For example, if we assume that the size of $R$-parity violating terms
are $O(\kappa_i)$ relative to the corresponding $R$-parity conserving
ones (which are obtained by replacing $\hat{H}_d$ with $\hat{L}_i$),
and that the size of the SUSY breaking parameters are typically of the
order of the electro-weak scale, then the above constraint is
consistent with the one obtained from the neutrino mass.

\subsection{Gravitino decay}

\begin{figure}[t]
  \begin{center}
    \includegraphics[scale=0.7]{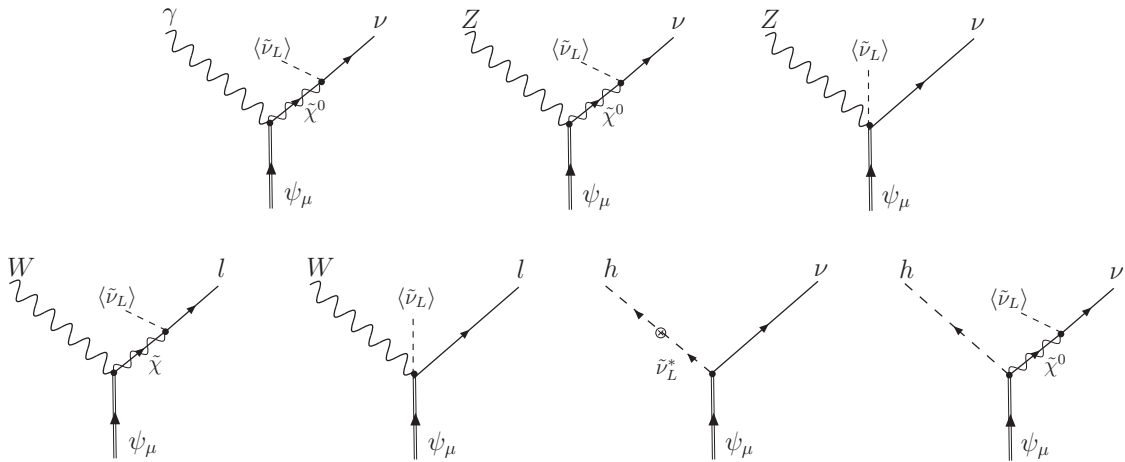} 
    \caption{\small Diagrams of gravitino decay.}
    \label{fig:decay}
  \end{center}
\end{figure}

In the case with RPV, gravitino LSP is no longer stable and decays to
standard-model particles with a finite lifetime
\cite{Takayama:2000uz}.  Here, we will take a closer look at the
gravitino decay.

In the present scenario, gravitino mainly decays in the two-body decay
processes shown in Fig.\ \ref{fig:decay}: $\psi_{\mu} \rightarrow
\gamma \nu_i$, $Z\nu_i$, $Wl_i$, and $h\nu_i$.  (Here and hereafter,
$\psi_\mu$ denotes the gravitino.)  Decay widths of each process are
given by\footnote
{In \cite{Ibarra:2007wg}, the coupling of the gravitino to the
  supercurrent of the slepton multiplet, which gives rise to the terms
  proportional to the functions $G$ and $H$, was neglected.
  Consequently, the decay rates of the gravitino into $W^\pm l^\mp$
  and $Z\nu$ are underestimated, resulting in enhanced branching ratio
  for the process $\psi_\mu\rightarrow\gamma\nu$.}
\begin{eqnarray}
  \Gamma_{\psi_{\mu}\rightarrow \gamma \nu_i}
  &=& \frac{1}{128 \pi} 
  \frac{\kappa_i^2 m_{3/2}^3}{M_{\rm Pl}^2} g_Z^2
  \theta_{\tilde{\gamma}}^2,
  \\
  \Gamma_{\psi_{\mu} \rightarrow Z \nu_i}
  &=& 
  \frac{\beta_Z}{128 \pi} 
  \frac{\kappa_i^2 m_{3/2}^3}{M_{\rm Pl}^2} 
  \biggl[
  g_Z^2 \theta_{\tilde{Z}}^2 F(m_{3/2},m_Z)
  + \frac{3v}{2m_{3/2}} 
  g_Z^2 \theta_{\tilde{Z}} G(m_{3/2},m_Z)
  \nonumber \\
  && 
  + \frac{1}{3} \beta_Z
%  \left(
%    1-\frac{m_Z^2}{m^2_{3/2}} 
%  \right)
  H(m_{3/2},m_Z)
  \biggr],
%  \left(
%    1-\frac{m_Z^2}{m^2_{3/2}}
%  \right),
  \\
  \Gamma_{\psi_{\mu} \rightarrow W l_i}
  &=& \frac{\beta_W}{64 \pi} \frac{\kappa^2_i m^3_{3/2}}{M^2_{\rm Pl}} 
  \biggl[
  g_2^2 \theta_{\tilde{W}}^2 F(m_{3/2},m_W)
  + \frac{3v}{2m_{3/2}} 
  g_2^2 \theta_{\tilde{W}} G(m_{3/2},m_W)
  \nonumber \\
  && 
  + \frac{1}{3} \beta_W H(m_{3/2},m_W)
  \biggr]
  \\
  \Gamma_{\psi \rightarrow h \nu_i}
  &=& 
  \frac{\beta_h^4}{384 \pi} \frac{\kappa_i^2 m^3_{3/2}}{M^2_{\rm Pl}} 
  \left(
  \frac{m^2_{\tilde{\nu}}}{m^2_{\tilde{\nu}}-m^2_h}
  + m_Z \sin \beta
  \sum_{\alpha} 
  \frac{c_{\tilde{H}^0_u \tilde{\chi}^0_{\alpha}}
    c_{\tilde{Z} \tilde{\chi}^0_{\alpha}}^*}{m_{\tilde{\chi}^0_{\alpha}}} 
  \right)^2,
  \label{Eq:widths}
\end{eqnarray}
where $M_{\rm Pl} \simeq 2.4 \times 10^{18}$ GeV is the reduced Planck
mass, $m_{3/2}$ is the gravitino mass,
\begin{eqnarray}
  \beta_X &\equiv& 1 - \frac{m_X^2}{m_{3/2}^2},
\end{eqnarray}
%with $m_X$ ($X=Z$, $W$, $h$) being the mass of the final-state gauge
%or Higgs boson, 
and the functions $F$, $G$, and $H$ are given by
\begin{eqnarray}
  F(m_{3/2},m_X) &=&
  1 - \frac{1}{3} \frac{m_X^2}{m^2_{3/2}}
  - \frac{1}{3} \frac{m_X^4}{m^4_{3/2}}
  - \frac{1}{3} \frac{m_X^6}{m^6_{3/2}},
  \\
  G(m_{3/2},m_X) &=&
  1 - \frac{1}{2} \frac{m_X^2}{m^2_{3/2}}
  - \frac{1}{2} \frac{m_X^4}{m^4_{3/2}},
  \\
  H(m_{3/2},m_X) &=&
  1 + 10 \frac{m_X^2}{m^2_{3/2}}
  + \frac{m_X^4}{m^4_{3/2}}.
\end{eqnarray}
In addition, we define
\begin{eqnarray}
  \theta_{\tilde{\gamma}} &\equiv& 
  v \sum_{\alpha=1}^4 
  \frac{c_{\tilde{\gamma} \tilde{\chi}^0_{\alpha}}
    c_{\tilde{Z} \tilde{\chi}^0_{\alpha}}^*}
       {m_{\tilde{\chi}^0_{\alpha}}},
  \label{theta_photino} \\
  \theta_{\tilde{Z}} &\equiv& 
  v \sum_{\alpha=1}^4 
  \frac{c_{\tilde{Z} \tilde{\chi}^0_{\alpha}}
    c_{\tilde{Z} \tilde{\chi}^0_{\alpha}}^*}
       {m_{\tilde{\chi}^0_{\alpha}}},
  \label{theta_zino} \\
  \theta_{\tilde{W}} &\equiv& 
  \frac{1}{2} v \sum_{\alpha=1}^2 
  \frac{c_{\tilde{W}^+ \tilde{\chi}^+_{\alpha}}
    c_{\tilde{W}^- \tilde{\chi}^-_{\alpha}}+ {\rm h.c.}}
       {m_{\tilde{\chi}^\pm_{\alpha}}},
  \label{theta_wino}
\end{eqnarray}
where $c_{\tilde{W}^{\pm} \tilde{\chi}^{\pm}_{\alpha}}$ is
the elements of unitary matrices which diagonalize the
mass matrix of charginos ${\cal M}_{\tilde{\chi}^\pm}$:
$\tilde{W}^{\pm}=\sum_{\alpha=1}^2 c_{\tilde{W}^{\pm}
\tilde{\chi}^{\pm}_{\alpha}} \tilde{\chi}^{\pm}_{\alpha}$
(i.e.,
$m_{\tilde{\chi}^\pm_\alpha}=\sum_{ij}
c_{i\tilde{\chi}^{\mp}_{\alpha}}
c_{j\tilde{\chi}^{\pm}_{\alpha}}[{\cal M}_{\tilde{\chi}^\pm}]_{ij}$).

\begin{figure}[t]
  \begin{center}
    \includegraphics[scale=1.2]{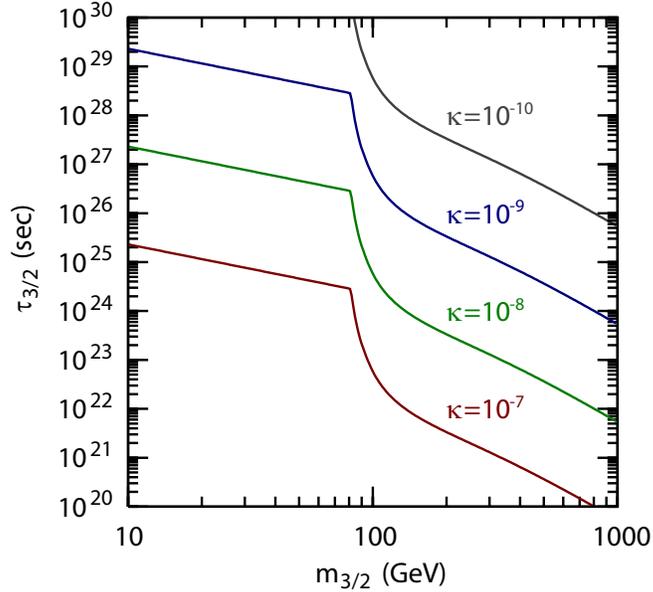} 
    \caption{\small Lifetime of gravitino as a function of gravitino
      mass.  Here, we take $\tan \beta=10, \ m_h=115 \ {\rm GeV}, \
      m_{\tilde{B}}=1.5m_{3/2}, \ m_{\tilde{\nu}}=2m_{3/2}$ under
      large Higgsino-mass limit, and assume GUT relation among gaugino
      masses.}
    \label{fig:tau}
  \end{center}
\end{figure}

\begin{figure}[t]
  \begin{center}
    \includegraphics[scale=1.2]{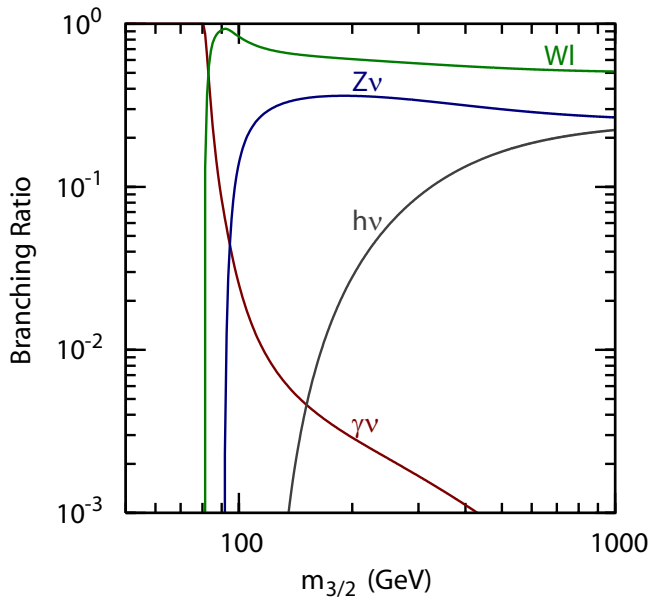} 
    \caption{\small Branching ratio for each decay mode.  Lines with
      the indices ``$Wl$,'' ``$Z\nu$,'' ``$h\nu$'' and ``$\gamma\nu$''
      show $Br(\psi_\mu\rightarrow W^+l^-)+Br(\psi_\mu\rightarrow
      W^-l^+)$, $Br(\psi_\mu\rightarrow Z\nu) +Br(\psi_\mu\rightarrow
      Z\bar{\nu})$, $Br(\psi_\mu\rightarrow h\nu)
      +Br(\psi_\mu\rightarrow h\bar{\nu})$, and
      $Br(\psi_\mu\rightarrow\gamma\nu)
      +Br(\psi_\mu\rightarrow\gamma\bar{\nu})$, respectively.
      (Summation over the generation index is implicit.)  Here, we
      take the MSSM parameters used in Fig.\ \ref{fig:tau} }
      \label{fig:br}
  \end{center}
\end{figure}

Lifetime of gravitino is determined by these two-body decay processes:
\begin{eqnarray}
\tau^{-1}_{3/2}
 = \Gamma_{3/2}
 = 2 \sum_{i=1}^3 
     \left[
          \Gamma_{\psi_{\mu}\rightarrow \gamma \nu_i}
        + \Gamma_{\psi_{\mu} \rightarrow Z \nu_i}
	+ \Gamma_{\psi_{\mu} \rightarrow W l_i}
	+ \Gamma_{\psi \rightarrow h \nu_i}
     \right],
\end{eqnarray}
where the factor of 2 is for CP-conjugated final states.  In Fig.\
\ref{fig:tau}, we plot $\tau_{3/2}$ as a function of $m_{3/2}$ in the
large Higgsino-mass limit.  Here, we take $\tan\beta =10$, $m_h=115 \
{\rm GeV}$, $m_{\tilde{\nu}}=2m_{3/2}$, $m_{\tilde{B}}=1.5m_{3/2}$,
and grand unified theory (GUT) relation among the gaugino masses are
assumed.  One can see that $\tau_{3/2}$ is much longer than the age of
the universe ($\simeq 4.3 \times 10^{17}\ {\rm sec}$) for weak-scale
gravitino mass when $\kappa\lesssim 10^{-7}$.  Thus, in such a
parameter region, most of gravitinos produced in the early universe
survive until the present epoch.

Even though the lifetime of the gravitino is long enough to realize
the gravitino dark matter scenario, it may be possible to observe the
decay of gravitino dark matter at present epoch.  In particular, high
energy photons and positrons are emitted in the decay processes as
well as in the following cascade decay processes.  To see this, in
Fig.\ \ref{fig:br}, we show branching ratio of each decay mode for
$\tan\beta =10$ in the large Higgsino mass limit.  (Notice that the
branching ratios are independent of $\kappa_i$.)

When $m_{3/2}\lesssim 80$ GeV, the decay mode
$\psi_{\mu}\rightarrow\gamma\nu$ dominates in total decay rate,
because the decay processes with the emission of the weak or Higgs
boson are kinematically blocked.  On the contrary, once the gravitino
becomes heavier than the weak bosons, the branching ratio for the
process $\psi_{\mu}\rightarrow\gamma\nu$ is suppressed.  This behavior
can be understood from the fact that the $\theta$-parameters defined
in Eqs.\ (\ref{theta_photino}) $-$ (\ref{theta_wino}), in particular,
$\theta_{\tilde{\gamma}}$, is suppressed when the neutralino and
chargino masses become much larger than the electro-weak scale.  

This fact has important implication in the study of the scenario using
cosmic rays.  When the gravitino is lighter than $\sim 100$ GeV or so,
line spectrum of the gamma ray may be a striking signal.  With larger
gravitino mass, on the contrary, it becomes difficult to observe the
gamma-ray line spectrum.  Even in that case, however, significant
amount of gamma rays with the energy of $O(1-100\ {\rm GeV})$ are
emitted in the cascade decays of $Z$, $W$, and $h$ bosons.  Therefore,
the continuous high energy gamma ray could be used as another
characteristic signal.  In addition to the gamma ray, high energy
positrons are also emitted, which is another interesting signal from
the decay of gravitino dark matter.

\section{Cosmic-Ray Fluxes: Formulae}
\label{sec:formalism}
\setcounter{equation}{0}

As we have discussed in the previous section, energetic gamma and
positron are produced if gravitino decays via $R$-parity violating
interactions.  If gravitino is dark matter of the universe, such decay
products can be a source of high energy cosmic rays.  In order to
discuss how well the scenario is tested by the use of cosmic ray, it
is necessary to formulate the calculation of the cosmic-ray fluxes
from the decay of gravitino.  In this section, we show how we
calculate the gamma-ray and positron fluxes from the gravitino decay.

\subsection{Gamma ray from the gravitino decay}

The total flux of the gamma ray from the decay of dark matter (i.e.,
gravitino) is calculated by the sum of two contributions:
\begin{eqnarray}
  \left[\frac{dJ_{\gamma}}{dE} \right]_{\rm DM} = 
  \left[\frac{dJ_{\gamma}}{dE} \right]_{\rm cosmo} + 
  \left[\frac{dJ_{\gamma}}{dE} \right]_{\rm halo},
\end{eqnarray}
where the first and second terms in the right-hand side are fluxes of
gamma ray from cosmological distance and that from the Milky Way halo,
respectively.  We discuss these contributions separately.

The flux of the gamma ray from cosmological distance is estimated as
\begin{eqnarray}
\left[E^2 \frac{dJ_{\gamma}}{dE} \right]_{\rm cosmo}
 = \frac{E^2}{m_{3/2} \tau_{3/2}} 
   \int_E^{\infty} dE^{\prime} G_{\gamma}(E,E^{\prime})  
   \frac{dN_{\gamma}(E^{\prime})}{dE^{\prime}}.
\end{eqnarray}
Here, the propagation function of gamma ray is given by
\begin{eqnarray}
  G_{\gamma}(E,E^{\prime})
  = \frac{c \rho_c \Omega_{3/2} }{4 \pi H_0 \Omega_M}
  \frac{1}{E} \left( \frac{E}{E^{\prime}} \right)^{3/2}
  \frac {1}{\sqrt{1+ \Omega_{\Lambda}/\Omega_M (E/E^{\prime})^3}},
\end{eqnarray}
where $c$ is the speed of light, $H_0$ is present Hubble expansion
rate, $\rho_c$ is critical density, and $\Omega_{3/2}\simeq
  0.1143h^{-2}$, $\Omega_M \simeq 0.1369h^{-2}$,
  $\Omega_{\Lambda}\simeq 0.721$ (with $h \simeq 0.701$) are
  density parameters of gravitino dark matter, total matter, and dark
  energy, respectively \cite{Hinshaw:2008kr}.  In addition,
$dN_{\gamma}/dE$ is the energy spectrum of gamma ray from the decay of
single gravitino, which can be given by the sum of contributions from
relevant decay modes:
\begin{eqnarray}
  \frac{dN_{\gamma}}{dE}
  &=& \frac{2}{\Gamma_{3/2}}
  \sum_{i=1}^{3}
  \Biggl(
  \Gamma_{\psi_{\mu} \rightarrow \gamma \nu_i}
  \left[ \frac{dN_{\gamma}}{dE} \right]_{\gamma \nu_i}
  + \Gamma_{\psi_{\mu} \rightarrow Z \nu_i}
  \left[ \frac{dN_{\gamma}}{dE} \right]_{Z \nu_i}
  \nonumber \\ &&
  + \Gamma_{\psi_{\mu} \rightarrow W l_i}
  \left[ \frac{dN_{\gamma}}{dE} \right]_{W l_i}
  + \Gamma_{\psi_{\mu} \rightarrow h \nu_i}
  \left[ \frac{dN_{\gamma}}{dE} \right]_{h \nu_i}
  \Biggr).
\end{eqnarray}
Here, $\left[ dN_{\gamma}/dE \right]_{\cdots}$ are energy
distributions for each decay modes.  Notice that the $dN_{\gamma}/dE$
is determined once the SUSY parameters are fixed, irrespective of the
cosmological scenario.  We calculate $dN_{\gamma}/dE$ by using PYTHIA
package \cite{pythia}.

The flux of the gamma ray from the Milky Way Galaxy halo is obtained
as \cite{Asaka:1998ju}
\begin{eqnarray}
  \left[E^2 \frac{dJ_{\gamma}}{dE} \right]_{\rm halo}
  = \frac{E^2}{m_{3/2} \tau_{3/2}} 
  \frac{1}{4 \pi}   \frac{dN_{\gamma}}{dE}
  \left\langle  
    \int_{\rm l.o.s.} \rho_{3/2}(\vec{l}) d \vec{l}
  \ \right\rangle_{\rm dir},
  \label{J(gamma)_halo}
\end{eqnarray}
where $\rho_{3/2}$ is the energy density of the gravitino in the Milky
Way halo.  In Eq.\ (\ref{J(gamma)_halo}), the integration should be
understood to extend over the line of sight (l.o.s.).  Thus, the
integration has an angular dependence on the direction of observation.
Here, $[E^2dJ_{\gamma}/dE]_{\rm halo}$ is given by averaging over the
direction, which is denoted as $\langle\cdots\rangle_{\rm dir}$.  In
the EGRET observation, the signal from the Galactic disc is excluded in
order to avoid the noise.  In order to compare our results with the
EGRET results, we also exclude the region within $\pm 10^{\circ}$
around the Galactic disk in averaging over the direction.

In order to perform the line-of-sight integration, the profile of
$\rho_{3/2}$, namely dark matter mass density profile $\rho_{\rm
  halo}$, should be given.  For our numerical analysis, we adopt
Navarro-Frenk-White (NFW) density profile \cite{NFW}:
\begin{eqnarray}
  \rho_{\rm halo}(r) =
  \frac{\rho_h}{r/r_c (1+r/r_c)^2},
  \label{eq:nfw}
\end{eqnarray}
where $r$ is the distance from the Galactic center, $\rho_h\simeq 0.33\
{\rm GeV}\ {\rm cm}^{-3}$, and $r_c \simeq 20$ kpc.  We have checked
that the dependence on the dark matter profile is negligible because, in
the calculation of the gamma-ray flux, we exclude the region around the
Galactic disc as we have mentioned.

\subsection{Positron from the gravitino decay}

Next, we discuss cosmic-ray positron from the gravitino decay.  If we
consider energetic positron propagating in galaxy, its trajectory is
twisted because of magnetic field.  With the expected strength of the
magnetic field, scale of gyro-radius of the trajectory is much smaller
than the size of the galaxy.  Furthermore, the magnetic field in the
galaxy is entangled.  Because of these, propagation of the positron in
the galaxy is expected to be well approximated as a random walk.

We use a diffusion model for the propagation of positron, in which
random walk is described by the following diffusion equation
\cite{Baltz:1998xv,Hisano:2005ec}:
\begin{eqnarray}
\frac{\partial  f_{e^+}(E,\vec{x})}{\partial t}
 = K(E) \nabla^2 f_{e^+}(E,\vec{x})
 + \frac{\partial}{\partial E}\left[ b(E) f_{e^+}(E,\vec{x}) \right]
 + Q(E,\vec{x}),
\end{eqnarray}
where $f_{e^+}(E,\vec{x})$ is the number density of positrons per unit
energy (with $E$ being the energy of positron), $K(E)$ is the
diffusion coefficient, $b(E)$ is the energy loss rate, and
$Q(E,\vec{x})$ is the positron source term.  As we mentioned above,
diffusion of injected positron is caused by the entangled magnetic
field in the galaxy.  On the other hand, the energy loss of the
positron is via Thomson and inverse Compton scatterings with Cosmic
Microwave Background and infrared gamma ray from stars or Synchrotron
radiation under the magnetic field.  The functions $K(E)$ and $b(E)$
can be determined so that the cosmic-ray Boron to Carbon ratio and
sub-Fe to Fe ratio are reproduced.  In our analysis, we use those
given in \cite{Baltz:1998xv}:
\begin{eqnarray}
  K(E) &=&
  3.3 \times 10^{27} \times \left[ 1.39 
    + \left( \frac{E}{1\ {\rm GeV}} \right)^{0.6}\right]
  {\rm cm^2\ sec^{-1}},
  \\
  b(E) &=& 
  10^{-16} \times 
  \left( \frac{E}{1\ {\rm GeV}} \right)^2 {\rm GeV sec^{-1}}.
\end{eqnarray}
Since the magnitude of the energy loss rate indicates that positron
loses its energy in the flight of less than a few kpc, the positron
flux from outside of our Milky Way Galaxy halo is negligible. Thus, in
the following discussion, we focus on the contribution of the positron
flux from the Milky Way Galaxy.  In addition, the positron source term
is given by the use of the positron injection rate and dark matter
distribution in the Milky Way Galaxy halo as\footnote
{The uncertainty on the positron flux from the effect of inhomogeneity
  in the local dark matter distribution is negligible
  \cite{BoostFactor}, which is quite contrast to the traditional case
  of dark matter annihilation.}
\begin{eqnarray}
  Q(E,\vec{x})= \frac{\rho_{\rm halo}(\vec{x})}{m_{3/2}}
  \frac{1}{\tau_{3/2}} \frac{dN_{e^+}}{dE},
\end{eqnarray}
where $dN_{e^+}/dE$ is energy distribution of positron from the decay
of single gravitino.  The explicit expression is given as
\begin{eqnarray}
  \frac{dN_{e^+}}{dE}
  &=& \frac{1}{\Gamma_{3/2}}
  \sum_{i=1}^{3}
  \Biggl(
   2 \Gamma_{\psi_{\mu} \rightarrow Z \nu_i}
  \left[ \frac{dN_{e^+}}{dE} \right]_{Z \nu_i}
  + \Gamma_{\psi_{\mu} \rightarrow W l_i}
  \left[ \frac{dN_{e^+}}{dE} \right]_{W^- l^+_i}
  \nonumber \\ &&
  + \Gamma_{\psi_{\mu} \rightarrow W l_i}
  \left[ \frac{dN_{e^+}}{dE} \right]_{W^+ l^-_i}  
  + 2 \Gamma_{\psi_{\mu} \rightarrow h \nu_i}
  \left[ \frac{dN_{e^+}}{dE} \right]_{h \nu_i}
  \Biggr),
\end{eqnarray}
where $[dN_{e^+}/dE]_{\cdots}$ is energy distribution for each decay
mode.  We calculate the energy distributions by the use of PYTHIA
package.  For density distribution, we adopt the same profile as the
previous section, namely NFW profile defined in Eq.\ (\ref{eq:nfw}).

We solve the diffusion equation in finite diffusion zone with boundary
condition.  Since the observed cosmic-ray positrons are considered to
be in equilibrium, we impose stability condition $\partial
f(E,\vec{x})/\partial t = 0$, and also free escape condition
$f(E,\vec{x})=0$ at the boundary.  The diffusion zone is usually
assumed as cylinder characterized with half-height $L$ and radius $R$.
Since positrons lose their energy after the flight of a few kpc or
less, the positron flux does not strongly depend on the choice of the
diffusion zone.  In our analysis, we take $L=4$ kpc and $R=20$ kpc.

Positron flux from the decay of gravitino dark matter is given by
\begin{eqnarray}
  \left[ \Phi_{e^+} (E) \right]_{\rm DM} \equiv
  \frac{dJ_{e^+}}{dE} =
  \frac{c}{4 \pi} f(E, \vec{R}_{\odot}), 
\end{eqnarray}
where $\vec{R}_{\odot}$ is the location of the solar system.  This
flux does not correspond exactly to the one observed on the top of the
atmosphere.  The flux is modified due to interaction with solar wind
and magneto-sphere. However, the modulation effect is not important
when the energy of a positron is above $10\ {\rm GeV}$. Furthermore,
the effect is highly suppressed in the positron fraction, which is
defined by the ratio of the positron flux to the sum of positron and
electron fluxes, i.e., $\Phi_{e^+}/(\Phi_{e^+}+\Phi_{e^-})$.

\subsection{Backgrounds}

In order to discuss high energy gamma ray and positron as signals from
the decay of dark-matter gravitino, it is important to understand
those cosmic rays from other sources.  (We call them backgrounds.) 

Cosmic gamma rays have various origins.  As we have mentioned, the
cosmic gamma rays can be divided into two parts by their origins;
Galactic and extragalactic parts.  Some part of the Galactic origins,
such as scattering processes (i.e., inverse Compton scattering and
bremsstrahlung) and pion decay, are known as probable sources of
cosmic gamma rays.  However, other Galactic origins, as well as the
extragalactic ones, have not been well understood.  In the study of
the gamma ray from the gravitino decay, it is important to understand
the behavior of the unidentified cosmic gamma ray (UCGR) from various
origins.\footnote
{The UCGR may have various origins.  Examples in astrophysics are
  contributions from galaxy clusters \cite{Ensslin:1996ep}, energetic
  particles in the shock waves associated with large-scale cosmological
  structure formation \cite{Loeb:2000na}, distant gamma-ray burst
  events, baryon-antibaryon annihilation \cite{Gao:1990bh}.  In
  addition, if we consider physics beyond the standard model, spectrum
  of UCGR may be affected by, for example, the evaporation of primordial
  black holes \cite{Hawking:1974rv}, the annihilation of WIMPs
  \cite{Jungman:1995df,Baltz:1998xv,AnnihilationPastWorks,Hooper:2004bq,Hisano:2005ec},
  or extragalactic IR and optical photon spectra \cite{Stecker:1998ux}.
  In our analysis, we only consider the decay of the gravitino as a
  particle-physics source of the UCGR, and do not consider other
  possibilities.}
However, theoretical calculation of such components of the cosmic
gamma ray is difficult, so we adopt more phenomenological approach
to extract the UCGR not originating from the gravitino decay.

Currently, the cosmic gamma ray flux has been measured by EGRET, and
various analysis have been performed to extract the UCGR from the
EGRET data.  The first intensive work was done in \cite{egret}, in
which it is concluded that the UCGR follows a power law as $E^2
dJ_{\gamma}/dE = 1.37 \times 10^{-6} (E/1 \ {\rm GeV})^{-0.1} ({\rm
  cm}^2 \ {\rm str} \ {\rm sec})^{-1} {\rm GeV}$.
Generally, however, there is a difficulty in removing the contribution
from the known scattering or decay processes in the Milky Way Galaxy
since the analysis depends on the Galactic model.
Recently, with an improved analysis in the estimation of the Galactic
contribution \cite{Moskalenko:1997gh,Strong:2004ry}, it has been
pointed out that the UCGR spectrum follows a power-law in the energy
range $E\lesssim 1$ GeV.  However, for $E\gtrsim 1$ GeV, a deviation
from the power-law behavior is reported.  As we will see, the effect
of the gravitino decay on the cosmic gamma ray becomes important for
the energy range of $E\gtrsim 0.1-1$ GeV.  Thus, in our analysis, we
assume that the UCGR in the lower energy range is only from
astrophysical origins (although many of them have not yet been well
understood).  In addition, we also assume that the spectrum of UCGR
from astrophysical origins follows a power law, and hence its behavior
can be extracted from the data in the sub-GeV region.  Since the gamma
ray from the gravitino becomes important above the energy of $0.1-1$
GeV, we use the observed data in the range of $0.05\ {\rm GeV}<E<
0.15\ {\rm GeV}$ to determine the background flux.  Assuming the
power-law behavior, we obtain the best-fit UCGR flux as
\begin{eqnarray}
  \left[E^2 \frac{dJ_{\gamma}}{dE} \right]_{\rm BG}
  \simeq 5.18 \times 10^{-7}\ 
  ({\rm cm}^2\ {\rm sec} \ {\rm str})^{-1} \ {\rm GeV} 
  \times \left( \frac{E}{{\rm GeV}} \right)^{-0.449}.
  \label{BG_gamma}
\end{eqnarray}
We use this spectrum as the background in the following analysis.
Then, the total gamma-ray spectrum is given by
\begin{eqnarray}
  \left[ \frac{dJ_{\gamma}}{dE} \right]_{\rm tot} = 
  \left[ \frac{dJ_{\gamma}}{dE} \right]_{\rm DM}
  + \left[ \frac{dJ_{\gamma}}{dE} \right]_{\rm BG}.
  \label{J_gamma}
\end{eqnarray}

Next, let us consider the background positron (and electron).  Cosmic
rays mainly consist of nuclei, electrons, and positrons.  Nuclei are
the dominant component of the cosmic rays and pouring to the earth
after it has drifted by interaction with interstellar matters in our
Galaxy.  As a consequence, secondary cosmic-ray electrons and
positrons are produced.  On the theoretical side, many simulations of
cosmic-ray electron and positron have been done by the use of
cosmic-ray propagation model \cite{Moskalenko:1997gh}.  In our study,
we adopt the following cosmic-ray electron and positron from
astrophysical processes \cite{Baltz:1998xv}:
\begin{eqnarray}
  \left[ \Phi_{e^-} \right]_{\rm prim}
  &=& 
  \frac{0.16 E_{\rm GeV}^{-1.1}}
  {1+11E_{\rm GeV}^{0.9}+3.2E_{\rm GeV}^{2.15}}
  ({\rm GeV \ cm}^2 \  {\rm sec \  str})^{-1},
  \\
  \left[ \Phi_{e^-} \right]_{\rm sec}
  &=& 
  \frac{0.70E_{\rm GeV}^{0.7}}
  {1+110E_{\rm GeV}^{1.5}+600E_{\rm GeV}^{2.9}+580E_{\rm GeV}^{4.2}}
  ({\rm GeV \ cm}^2 \ {\rm sec \  str})^{-1},
  \\
  \left[ \Phi_{e^+} \right]_{\rm sec}
  &=& \frac{4.5E_{\rm GeV}^{0.7}}
  {1+650E_{\rm GeV}^{2.3}+1500E_{\rm GeV}^{4.2}}
  ({\rm GeV \ cm}^2 \ {\rm sec \  str})^{-1},
  \label{BG_e+}
\end{eqnarray}
where $E_{\rm GeV}$ is the energy of electron or positron in units of
GeV.  We note here that concerning the backgrounds, the secondary
electron accounts for about 10\ \% of the total electron flux while
the positron flux is dominated by the secondary one.

With these backgrounds, the total fluxes of the electron and positron
are obtained as
\begin{eqnarray}
  \left[ \Phi_{e^+} \right]_{\rm tot}
  &=& \left[ \Phi_{e^+} \right]_{\rm DM}
  + \left[ \Phi_{e^+} \right]_{\rm sec},
  \label{Phi_e+}
  \\
  \left[ \Phi_{e^-} \right]_{\rm tot}
  &=& 
  \left[ \Phi_{e^-} \right]_{\rm DM}
  + \left[ \Phi_{e^-} \right]_{\rm prim}
  + \left[ \Phi_{e^-} \right]_{\rm sec}.
  \label{Phi_e-}
\end{eqnarray}

Importantly, the above background in cosmic-ray positron flux well
agrees with HEAT observation in the energy range $E\lesssim 10$ GeV.
However, for $E \gtrsim 10$ GeV, an excess of the positron flux is
seen in the HEAT data \cite{heat}.  In the following section, we show
that such an excess may be due to the decay of gravitino.\footnote
{For other possibilities, see 
\cite{Jungman:1995df,Baltz:1998xv,AnnihilationPastWorks,Hisano:2005ec}.}

\section{Numerical Results}
\label{sec:results}
\setcounter{equation}{0}

\subsection{Cosmic rays from the gravitino decay}

Now we are at the position to present our numerical results. In this
subsection, we show fluxes of cosmic-ray gamma and positron from the
gravitino decay to discuss their behaviors.

\begin{figure}[t]
  \begin{minipage}{0.5\hsize}
    \begin{center}
      \includegraphics[width=70mm]{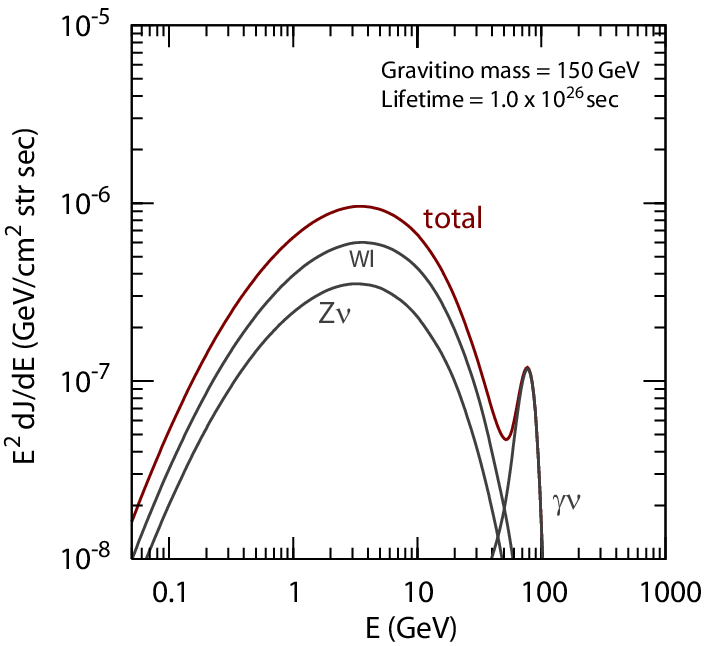}
    \end{center}
  \end{minipage}
  \begin{minipage}{0.5\hsize}
    \begin{center}
      \includegraphics[width=70mm]{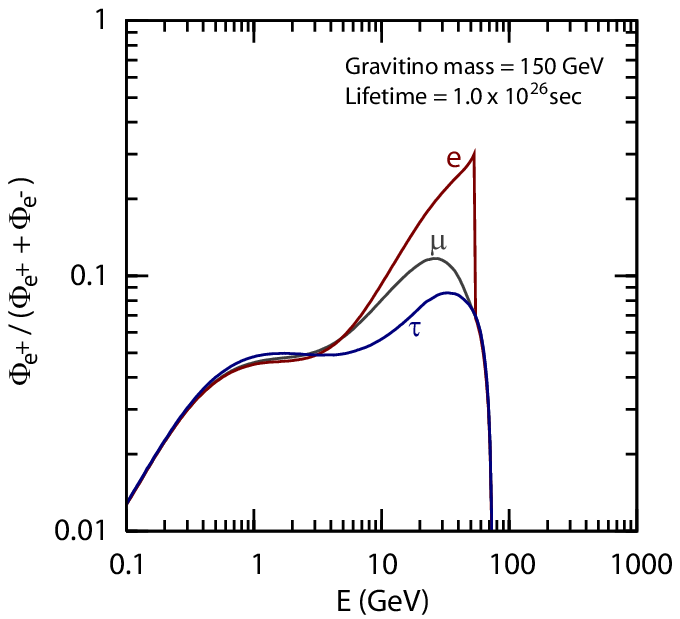}
    \end{center}
  \end{minipage}
  \caption{\small Left: Gamma-ray flux from gravitino decay $[E^2
    dJ_{\gamma}/dE]_{\rm DM}$(``total'').  Lines with ``$\gamma\nu$,''
    ``$Z\nu$,'' and ``$Wl$'' are contributions from each decay mode.
    Right: Positron fraction $[\Phi_{e^+}]_{\rm DM}/([\Phi_{e^+}]_{\rm
      tot}+[\Phi_{e^-}]_{\rm tot})$.  Lines with ``$e$,'' ``$\mu$,''
    and ``$\tau$'' show the results for the case that the gravitino
    mainly decays to first, second, and third generation leptons,
    respectively.  For both of the figures, we take $m_{3/2}=150~{\rm
      GeV}$, $\tau_{3/2}=1.0\times 10^{26}~{\rm sec}$, and the MSSM
    parameters used in Fig.\ \ref{fig:tau}.}
  \label{fig:sampleflux}
\end{figure}

First, we discuss the gamma-ray flux.  The flux for $m_{3/2}=150\ {\rm
  GeV}$ is shown on the left in Fig.\ \ref{fig:sampleflux}.  (Here,
the lifetime of $\tau_{3/2}=10^{26}\ {\rm sec}$ is used.)  We found
that the dependence of the gamma-ray flux on the flavor of the primary
lepton is negligible.  (Here and hereafter, for the calculation of the
gamma-ray flux, we assume that the gravitino mainly decays into
third-generation leptons and gauge or Higgs bosons.)  In calculating
the gamma-ray flux, we adopt the energy resolution of 15\ \%,
following EGRET \cite{egret}.  In the figures, we also show the
contributions of individual decay modes.  (In this case, the
contribution of the decay mode into $h\nu$ is very small.)

As one can see, continuous spectrum is obtained from the decay modes
into weak boson (or Higgs boson) and lepton, while a relatively steep
peak is also obtained at $E=\frac{1}{2}m_{3/2}$ due to the
monochromatic gamma emission via $\psi_{\mu} \rightarrow\gamma\nu$.
With larger gravitino mass, $Br(\psi_{\mu} \rightarrow\gamma\nu)$ is
suppressed, and hence the peak at $E=\frac{1}{2}m_{3/2}$ becomes less
significant.  Thus, when the gravitino mass becomes much larger than
the masses of weak bosons, it will become difficult to find such a
line spectrum.

The positron fraction is shown on the right in Fig.\
\ref{fig:sampleflux} for the cases where the gravitino dominantly
decays into first, second, and third generation leptons.  Since the
energy of positron becomes smaller during the propagation in the
Galaxy, the positron spectrum has the upper end point at
$\sim\frac{1}{2}m_{3/2}$.  In particular, when the gravitino can
directly decay into positron, the end point becomes a steep edge; such
an edge can be a striking signal of the decaying dark matter if
observed.  Even in other cases, the positron spectrum has a peak just
below the upper end point.  This may also provide an interesting
signal in the observed positron flux.

As one can see, the fluxes of the gamma-ray and positon in the present
scenario are most enhanced for the energy of $O(1-100\ {\rm GeV})$.  In
addition, the fluxes are significantly suppressed for the energy smaller
than $\sim 1\ {\rm GeV}$.  Thus, the present scenario is not constrained
from the observations in such a low-energy region, like the observations
of the fluxes of sub-GeV gamma rays.

So far, we have shown results only for the case of
$\tau_{3/2}=10^{26}\ {\rm sec}$.  Fluxes with other values of the
lifetime can be easily obtained from the above results since the
cosmic-ray fluxes from the gravitino decay is inversely proportional
to $\tau_{3/2}$.  One important point is that, when
$\tau_{3/2}=O(10^{26}\ {\rm sec})$, both the gamma-ray and positron
fluxes from the gravitino decay become comparable to the background
fluxes discussed in the previous section for $E\sim 1 - 100\ {\rm
  GeV}$.  Thus, with such a lifetime, we may be able to see the signal
of the decay of gravitino dark matter in the spectrum of cosmic rays.
The following subsections are devoted to the discussion of such an
issue.

\subsection{Implications to present observations}

As discussed in the previous section, some anomalies are indicated
both in the gamma-ray spectrum observed by EGRET and the positron
fraction observed by HEAT.  In this subsection, we show that these
anomalies may be simultaneously explained by a single scenario, the
gravitino dark matter scenario with RPV.

\begin{figure}[t]
  \begin{minipage}{0.5\hsize}
    \begin{center}
      \includegraphics[width=70mm]{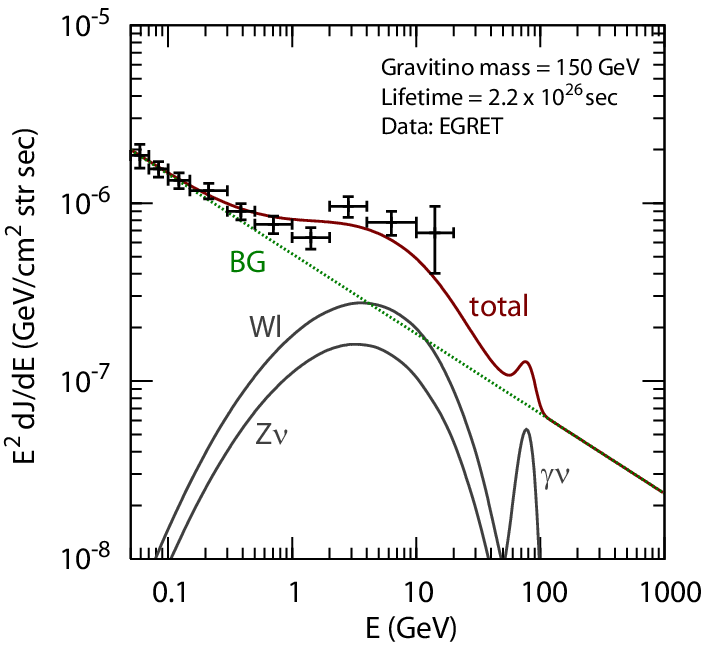}
    \end{center}
  \end{minipage}
  \begin{minipage}{0.5\hsize}
    \begin{center}
      \includegraphics[width=70mm]{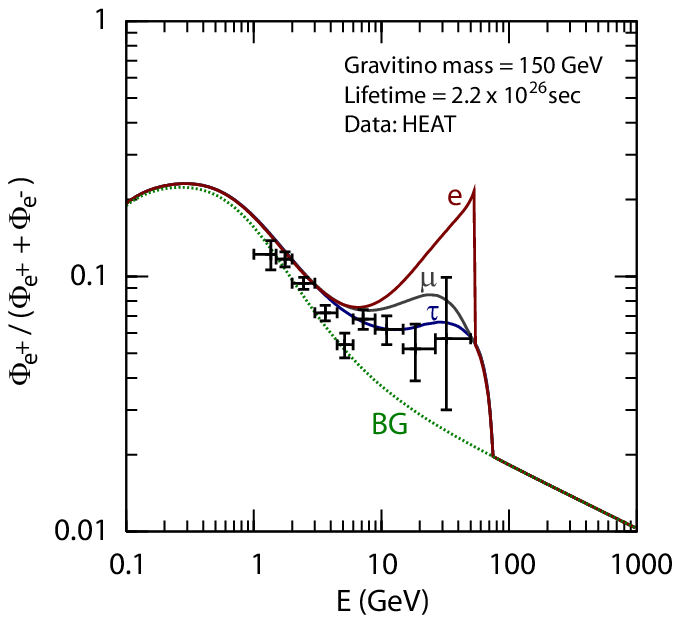}
    \end{center}
  \end{minipage}
  \caption{\small Gamma-ray flux (left figure) and positron fraction
    (right figure).  Here, we take $m_{3/2}=150~{\rm GeV},\
    \tau_{3/2}=2.2 \times 10^{26}~{\rm sec}$, and MSSM parameters as
    Fig.\ \ref{fig:tau}.}
 \label{fig:Mg150}
\end{figure}

In Fig.\ \ref{fig:Mg150}, we show the total gamma-ray flux and the
positron fraction for $m_{3/2}=150\ {\rm GeV}$ and
$\tau_{3/2}=2.2\times 10^{26}\ {\rm sec}$.  (We have used the best fit
value of $\tau_{3/2}$ for the EGRET data.)  Here, we consider simple
cases where the gravitino decays only into one of the three lepton
flavors; only one of the $\langle\tilde{\nu}_{i}\rangle$ ($i=1-3$) is
non-vanishing while the others are set to be zero.  (For gamma ray,
the total flux is independent of generation indices as we have
mentioned in the previous subsection.)  With the lifetime adopted, the
gamma-ray flux and the positron fraction both significantly deviate
from the background.  In gamma-ray flux (Fig.\ \ref{fig:Mg150}, left),
one can see that the continuous spectrum originating from the
processes $\psi_{\mu} \rightarrow Wl$ and $Z\nu$ gives a good
agreement with EGRET data for $E \sim 1-10 {\rm GeV}$.  In the
positron fraction (Fig.\ \ref{fig:Mg150}, right), the results indicate
that clear signal can be seen in the energy region $E\gtrsim$10 GeV
over the background for all the three cases.   In the same figure,
we also show the observational data of EGRET or HEAT.  As one can see,
agreements between the theoretical predictions and observations are
improved both for EGRET and HEAT.

\begin{figure}
  \begin{minipage}{0.5\hsize}
    \begin{center}
      \includegraphics[width=70mm]{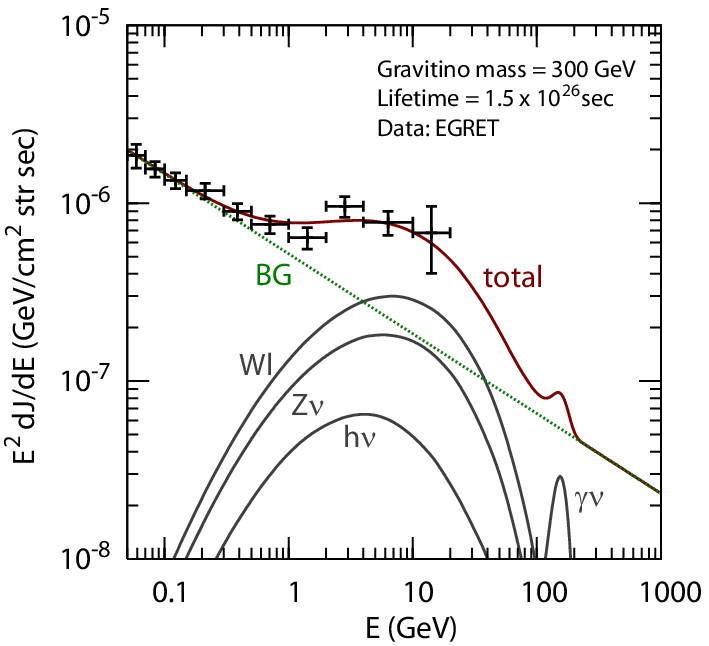}
    \end{center}
  \end{minipage}
  \begin{minipage}{0.5\hsize}
    \begin{center}
      \includegraphics[width=70mm]{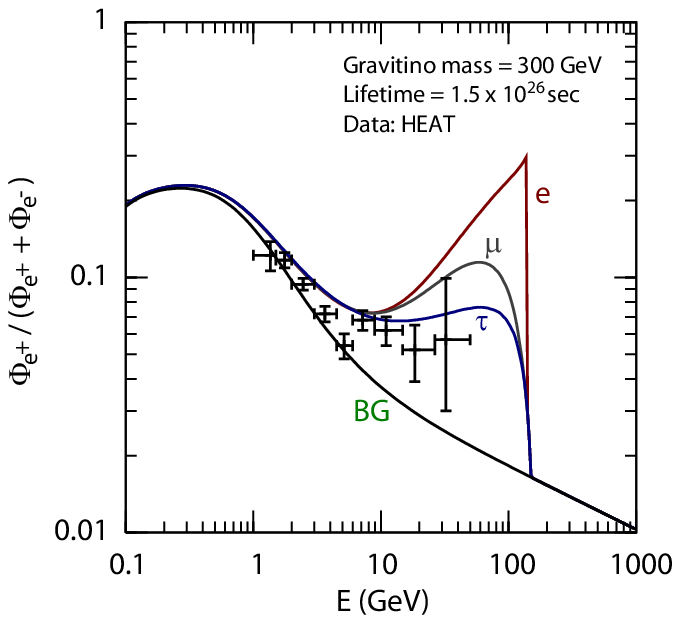}
    \end{center}
  \end{minipage}
  \caption{\small Same as Fig.\ \ref{fig:Mg150}, except for
    $m_{3/2}=300\ {\rm GeV}$ and $\tau_{3/2}=1.5 \times 10^{26}\ {\rm
      sec}$.}
  \label{fig:Mg300}
\end{figure}

\begin{figure}
  \begin{minipage}{0.5\hsize}
    \begin{center}
      \includegraphics[width=70mm]{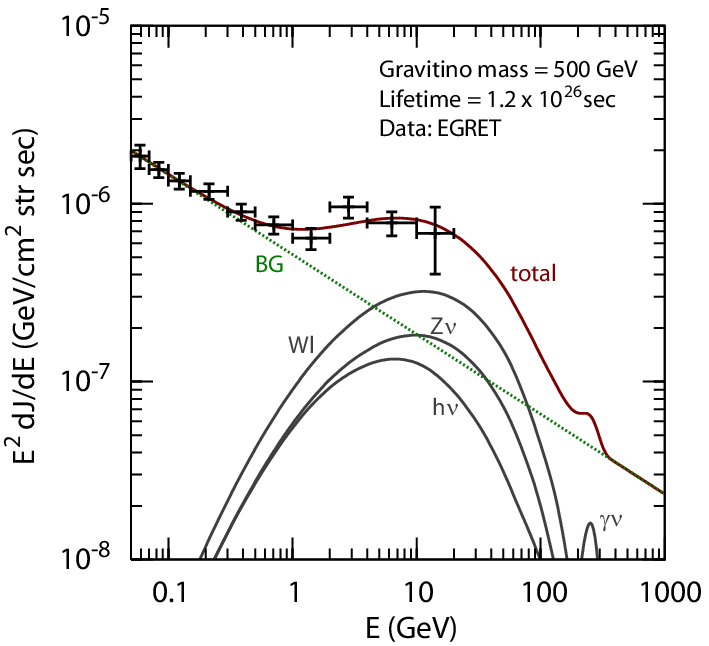}
    \end{center}
  \end{minipage}
  \begin{minipage}{0.5\hsize}
    \begin{center}
      \includegraphics[width=70mm]{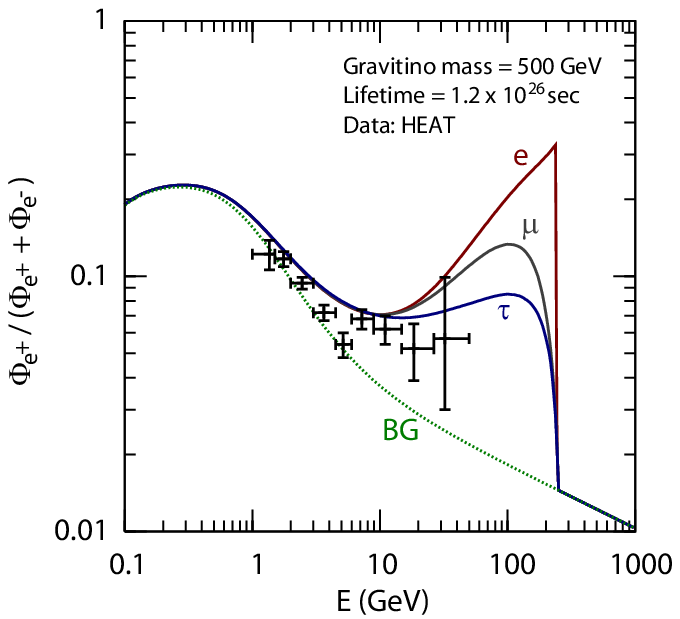}
    \end{center}
  \end{minipage}
  \caption{\small Same as Fig.\ \ref{fig:Mg150}, except for
    $m_{3/2}=500\ {\rm GeV}$ and $\tau_{3/2}=1.2 \times 10^{26}\ {\rm
      sec}$.}
 \label{fig:Mg500}
\end{figure}

Results for other values of the gravitino mass (and lifetime) are
shown in Figs.\ \ref{fig:Mg300} and \ref{fig:Mg500}.  We can see that
the suggested anomalies in the gamma-ray and positron fluxes may be
explained in a wide range of the gravitino mass.

\begin{figure}
  \begin{center}
    \includegraphics[scale=1.2]{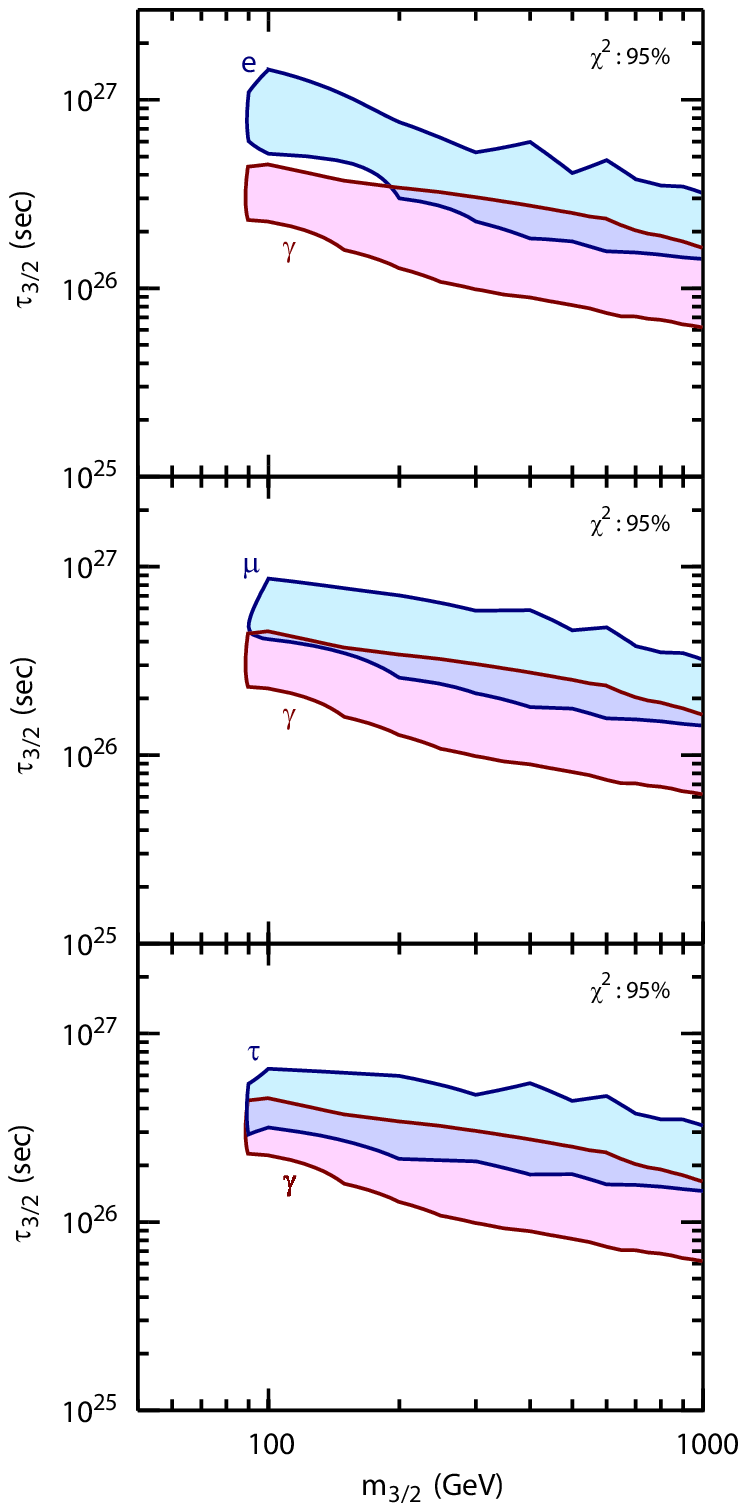} 
    \caption{\small 95\% C.L. allowed regions on $m_{3/2}$ vs.\
      $\tau_{3/2}$ plane; the shaded regions are the allowed regions.
      The region with ``$\gamma$'' is for the EGRET data, while the
      region with ``$e$'' (``$\mu$,'' ``$\tau$'') is for HEAT data for
      the case where the gravitino decays into first- (second-,
      third-) generation lepton.}
    \label{fig:chisqcontour}
  \end{center}
\end{figure}

In order to see the preferred parameter region in the light of EGRET
result, we calculate the $\chi^2$ variable as a function of $m_{3/2}$
and $\tau_{3/2}$:
\begin{eqnarray}
  \chi^2
  = \sum_{i=1}^N 
  \frac{(x_{{\rm th},i}-x_{{\rm obs},i})^2}{\sigma^2_{{\rm obs},i}},
\end{eqnarray}
where $x_{{\rm th},i}$ is the theoretically calculated flux in $i$-th
bin, which is calculated with Eq.\ (\ref{J_gamma}), $x_{{\rm obs},i}$
is the observed flux, and $\sigma_{{\rm obs},i}$ is the error of
$x_{{\rm obs},i}$.  In addition, $N$ is number of bins; $N=10$ for
EGRET.  In Fig.\ \ref{fig:chisqcontour}, we show the region with
$\chi^2<18.3$ on the $m_{3/2}$ vs.\ $\tau_{3/2}$ plane, which is 95\
\% C.L. allowed region.  As one can see, the present scenario could
well explain the EGRET anomaly in a wide parameter region, $10^{26}\
{\rm sec}\lesssim \tau_{3/2}\lesssim 10^{27}\ {\rm sec}$ and
$m_{3/2}\gtrsim 90\ {\rm GeV}$.  From Fig.\ \ref{fig:tau}, it can be
seen that $10^{-10} \lesssim \kappa_i \lesssim 10^{-8}$ is favored.
In Fig.\ \ref{fig:chisqcontour}, we also show the parameter region
which is consistent with the HEAT data ($N=9$) at 95\ \% C.L. 
(i.e., $\chi^2< 16.9$).

 As one
can see, the present scenario can simultaneously explain the observed
gamma and positron fluxes.

\subsection{Future prospects}

In the previous subsection, we have shown that the gamma-ray and
positron fluxes from decaying gravitinos can successfully explain the
results of the past observations.  As we have seen, however, the
energy ranges of the past observations are limited up to $O(10\ {\rm
  GeV})$ although the signal from the gravitino decay may
significantly affect the cosmic-ray spectra up to the energy of
$O(100\ {\rm GeV})$.  In addition, it is also true that the
uncertainties of the cosmic-ray spectra observed by the past
observations are relatively large at the energy range of $E\sim O(10\
{\rm GeV})$.  Thus, it is desirable to test the scenario of
gravitino dark matter with RPV with better observations.

Fortunately, in the near future, new observations of cosmic rays,
Gamma-ray Large Area Space Telescope (GLAST) and Payload for
Antimatter Matter Exploration and Light nuclei Astrophysics (PAMELA),
are expected to provide results of new measurements of the cosmic-ray
fluxes.  These experiments are designed to detect cosmic rays with
energy up to a few hundreds GeV.  Thus, they will give us better test
of the scenario.  Since GLAST (PAMELA) has better energy range and
resolution than EGRET (HEAT) in the measurement of gamma-ray
(positron) flux, they should confirm the anomalies if they really
exist.

Even if the fluxes of the cosmic rays are smaller than the best-fit
value of those observed by EGRET and HEAT, we still have a chance to
see signals from the decay of dark-matter gravitino.  To see expected
constraints on the parameter space, we calculate the expectation value
of the $\chi^2$ variable defined as
\begin{eqnarray}
  \langle \chi^2 \rangle
  = \Bigl\langle 
  \sum_{i=1}^{N} \frac{(N_{{\rm th},i}-N_{{\rm BG},i})^2}
  {\sigma^2_{{\rm BG},i}}
  \Bigr\rangle ,
\end{eqnarray}
where $N$ is the number of bins.  We use 50 bins to estimate the
expected sensitivities of GLAST and PAMELA.  With the use of GLAST and
PAMELA instrument parameters \cite{glast,pamela,Hooper:2004bq}, we
define $i$-th bin as $E_{\gamma/e^+,i}^{\rm (min)}\leq
E_{\gamma/e^+,i}< E_{\gamma/e^+,i}^{\rm (max)}$, where
\begin{eqnarray}
  E_{\gamma,i}^{\rm (min)} &=& 0.02\ {\rm GeV} \times 
  \left(\frac{300~{\rm GeV}}{0.02~{\rm GeV}} \right)^{\frac{i-1}{50}},
  \\
  E_{e^+,i}^{(min)} &=& 0.05\ {\rm GeV} \times 
  \left(\frac{270~{\rm GeV}}{0.05~{\rm GeV}} \right)^{\frac{i-1}{50}},
\end{eqnarray}
and $E_{\gamma/e^+,i}^{\rm (max)}=E_{\gamma/e^+,i+1}^{\rm (min)}$.  In
addition, $N_{{\rm th},i}$ is the number of theoretically calculated
events in $i$-th bin, which is the expected number of events in GLAST
or PAMELA experiment.  Notice that $N_{{\rm th},i}$ is calculated by
using Eq.\ (\ref{J_gamma}) or Eq.\ (\ref{Phi_e+}).  Furthermore,
$N_{{\rm BG},i}$ is the number of background events.  Here, we assume
that the number of background events will be well understood in the
future observations by using the date in low-energy range; in our
following study, we use Eqs.\ (\ref{BG_gamma}) and (\ref{BG_e+}).
Then, we obtain
\begin{eqnarray}
  N_{{\rm th},i}
  &=& \left[ \frac {dJ_{\gamma}}{dE} (E_{\gamma,i}) \right]_{\rm tot} 
  \Delta E_{\gamma,i} T S,
  \\
  N_{{\rm BG},i}
  &=& \left[ \frac{dJ_{\gamma}}{dE} (E_{\gamma,i}) \right]_{\rm BG}
  \Delta E_{\gamma,i} T S,
\end{eqnarray}
for gamma-ray flux, and 
\begin{eqnarray}
  N_{{\rm th},i}
  &=& 
  \left[ \Phi_{e^+}(E_{e^+,i}) \right]_{\rm tot} \Delta E_{e^+,i} T S,
  \\
  N_{{\rm BG},i}
  &=& 
  \left[ \Phi_{e^+}(E_{e^+,i}) \right]_{\rm sec}\Delta E_{e^+,i} T S,
\end{eqnarray}
for cosmic-ray positron flux.  Here, $T$ and $S$ are exposure time and
acceptance, respectively, and $\sigma_{{\rm BG},i}$ is the error of
the number of backgrounds, which we take $\sigma_{{\rm
    BG},i}=\sqrt{N_{{\rm BG},i}}$ assuming that the error is dominated
by statistics.  In our analysis, we take $TS = 10^{10}$ and $10^{8}\
{\rm cm^2\ sec\ str}$ for GLAST and PAMELA, respectively.  In
addition, $\Delta E_{\gamma/e^+,i}=E_{\gamma/e^+,i}^{\rm
  (max)}-E_{\gamma/e^+,i}^{\rm (min)}$ is the width of the $i$-th bin.

\begin{figure}[t]
 \begin{center}
  \includegraphics[scale=1.2]{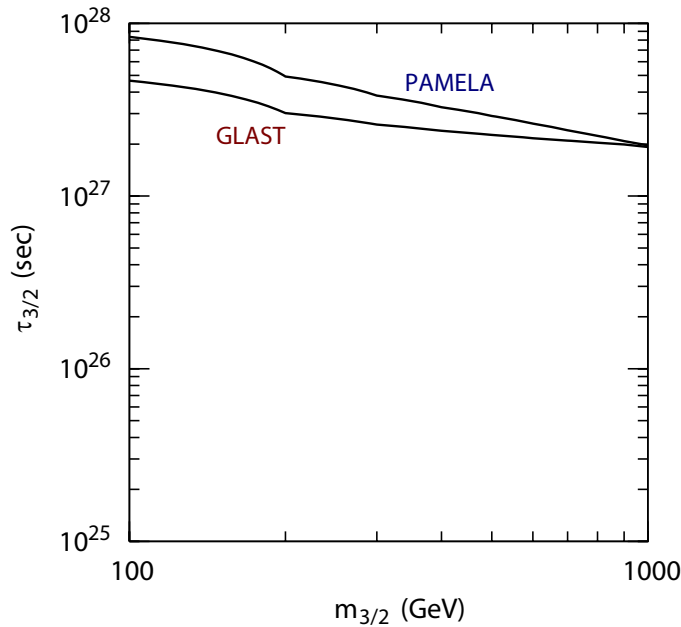} 
  \caption{\small Contour plot of $\langle \chi^2 \rangle$ on
    $m_{3/2}$ vs.\ $\tau_{3/2}$ plane for 95\% probability to detect
    gamma-ray/positron signal from gravitino decay in
    GLAST/PAMELA. } \label{fig:disGlastPamela}
 \end{center}
\end{figure}

Expected constraints on the $m_{3/2}$ vs.\ $\tau_{3/2}$ plane is shown
in Fig.\ \ref{fig:disGlastPamela}.  For positron flux, we consider the
case that the primary lepton emitted by the gravitino decay is $\tau$
or $\nu_\tau$.  In Fig.\ \ref{fig:disGlastPamela}, we show contours of
$\langle\chi^2\rangle=67.5$, corresponding to 95\ \% level of
detectability.  From the figure, one can see that GLAST and
  PAMELA have sensitivities for the case with the lifetime of
  $O(10^{27}\ {\rm sec})$, and hence the signal from decaying gravitino
  can be observed in both observations in wide range of the parameter
  space.  We have seen that, in order to explain the EGRET and HEAT
  anomalies, $\tau_{3/2}$ is required to be $\sim 10^{26}\ {\rm sec}$,
  which is order of magnitude shorter then the reach of GLAST and
  PAMELA experiments.  Thus, even if EGRET or HEAT anomaly somehow
  dissapears, GLAST or PAMELA still has a chance to find some
  high-energy-cosmic-ray signals from the gravitino dark matter
  scenario with $R$-partity violation.\footnote
  {Recently, the positron flux measured by PAMELA has been reported
    \cite{PAMELA08}.  Importantly, the PAMELA result is still
    preliminary, and the positron flux is shown only up to the energy
    of $10\ {\rm GeV}$.  The new result is inconsistent with the
    previously calculated background flux, so the understanding of the
    background will become very important if the present result will
    be confirmed with more data.  Thus, even though the PAMELA result
    is also inconsistent with the HEAT result, we think that it is
    still premature to discuss its implication.}

\section{Conclusions and Discussion}
\label{sec:discussion}
\setcounter{equation}{0}

In this paper, we have discussed the high energy cosmic rays from the
decay of gravitino dark matter in $R$-parity violated supersymmetric
model.  Here, we have considered the case that the gravitino is the
LSP, and that RPV originates from the $L_iH_u$-type operators.  In
such a model, the gravitino dominantly decays into a gauge boson
($W^\pm$, $Z$, or $\gamma$) and a lepton with extremely long lifetime.
In particular, if RPV interactions are weak enough, the lifetime of
the gravitino becomes much longer than the present age of the
universe.  Consequently, if the right amount of the gravitino is
produced in the early universe, gravitino can be dark matter even if
the $R$-parity is not conserved.  In addition, in such a scenario, the
MSSM-LSP, which is assumed to be the NLSP, can decay with a lifetime
shorter than $\sim 1\ {\rm sec}$ via RPV operators.  Then, the serious
effects on the light element abundances by the decay of the NLSP,
which gives one of the most stringent constraint on the scenario of
gravitino dark matter, can be avoided.

We have studied the gamma-ray and positron fluxes from the decay of
the gravitino dark matter.  In our analysis, we have calculated the
decay rate and branching ratios of the gravitino, taking account of
all the relevant operators.  Then, the energy spectrum of the primary
decay products (i.e., gauge bosons, leptons, and partons) are
calculated.  Decay and hadronization processes of those primary
particles have been treated by using PYTHIA package to accurately
calculate the primary fluxes of gamma ray and positron by the decay of
gravitino.  Then, by solving the propagation equations, we have
obtained the fluxes of gamma ray and positron in the cosmic ray.

One of our important results is that the anomalies observed by the
EGRET and HEAT experiments can be simultaneously explained in this
scenario if the lifetime of the gravitino is $O(10^{26}\ {\rm sec})$.
This conclusion holds for a wide range of the gravitino mass; as far
as $m_{3/2}$ is larger than $\sim 80\ {\rm GeV}$ so that the gravitino
can decay into weak boson(s), such a scenario works.  In addition,
with PAMELA and GLAST, more accurate test of the scenario will become
possible; they will cover the parameter region of
$\tau_{3/2}\lesssim 10^{27-28}\ {\rm sec}$.

We comment here that, in the present scenario, the MSSM-LSP decays via
RPV interactions, which may affect the LHC phenomenology.  If the
EGRET and HEAT anomalies are due to the decay of the gravitino dark
matter, the lifetime of the MSSM-LSP is estimated to be $O(10^{-5}\
{\rm sec})$.  This fact may have an impact on the LHC experiment.  The
typical decay length of the MSSM-LSP is $O(10^3\ {\rm m})$, which is
much longer than the size of the detector.  Thus, most of the
MSSM-LSPs produced at the LHC experiment escape from the detector
before the decay.  If the lightest neutralino is the LSP, the collider
signatures are the same as the conventional signatures with the
neutralino LSP.  However, in the present scenario, the MSSM-LSP may be
charged (or even colored).  In such a case, we will observe a heavy
charged particle as high $p_T$ track at the LHC.  In addition, $0.1-1\
\%$ of the produced MSSM-LSP may decay in the detector, resulting in
isolated vertices from the interaction point.  These are very exotic
signals which are not expected in the conventional model of
supersymmetry.

So far, we have considered the case where the gravitino is dark
matter.  However, with other candidates for dark matter, it may be
possible to simultaneously explain the excesses of the gamma-ray and
positron fluxes observed by EGRET and HEAT experiments.  In
particular, if dark matter decays mainly into the weak bosons with the
lifetime of $\sim 10^{26}\ {\rm sec}$, this can be the case.  One of
the examples may be the lightest neutralino.  With RPV, the lightest
neutralino decays even if it is the LSP; with the RPV operator given
in Eq.\ (\ref{L_RPV}), for example, the lifetime of the neutralino LSP
becomes $\sim 10^{26}\ {\rm sec}$ when $\kappa\sim 10^{-25}$, assuming
that the lightest neutralino is Bino-like and that the mass of the
lightest neutralino is $\sim 100\ {\rm GeV}$.

\noindent {\it Note added:} While finalizing this paper, we found
\cite{Ibarra:2008qg} which also studies high energy cosmic rays from the
decay of the gravitino dark matter.  In this paper, antiproton flux is
also calculated, and it is discussed that the antiproton flux may become
too large if we explain the EGRET and HEAT anomalies in the present
scenario.  However, it is also mentioned that the predicted antiproton
flux suffers from large uncertainties.  Thus, according to
\cite{Ibarra:2008qg}, taking into account the uncertainties, the
antiproton flux can be consistent with the observations even if we
explain the EGRET and HEAT anomalies. In addition, in the study of the
gravitino decay in \cite{Ibarra:2008qg}, effects of the coupling of the
gravitino to the supercurrent of the slepton multiplet have not been
taken into account.

\noindent
{\it Acknowledgments:}
This work was supported in part by Research Fellowships of the Japan
Society for the Promotion of Science for Young Scientists (K.I.), and
by the Grant-in-Aid for Scientific Research from the Ministry of
Education, Science, Sports, and Culture of Japan, No.\ 19540255
(T.M.).


\begin{thebibliography}{99}

\bibitem{Buchmuller:2007ui}
  W.~Buchmuller, L.~Covi, K.~Hamaguchi, A.~Ibarra and T.~Yanagida,
  JHEP {\bf 0703}, 037 (2007).
  %%CITATION = JHEPA,0703,037;%%

\bibitem{Fukugita:1986hr}
  M.~Fukugita and T.~Yanagida,
  Phys.\ Lett.\  B {\bf 174}, 45 (1986).
  %%CITATION = PHLTA,B174,45;%%

\bibitem{Ibarra:2007wg}
  A.~Ibarra and D.~Tran,
  Phys.\ Rev.\ Lett.\  {\bf 100}, 061301 (2008).
  %%CITATION = PRLTA,100,061301;%%

\bibitem{pythia}
  T.~Sjostrand, S.~Mrenna and P.~Skands,
  %``PYTHIA 6.4 physics and manual,''
  JHEP {\bf 0605}, 026 (2006)
  %%CITATION = JHEPA,0605,026;%%

\bibitem{egret}
  P.~Sreekumar {\it et al.}  [EGRET Collaboration],
  %``EGRET observations of the extragalactic gamma ray emission,''
  Astrophys.\ J.\  {\bf 494}, 523 (1998).
  %%CITATION = ASJOA,494,523;%%

\bibitem{heat}
  S.~W.~Barwick {\it et al.}  [HEAT Collaboration],
  %``Measurements of the cosmic-ray positron fraction from 1-GeV to 50-GeV,''
  Astrophys.\ J.\  {\bf 482}, L191 (1997).
  %%CITATION = ASJOA,482,L191;%%

\bibitem{Jungman:1995df}
  G.~Jungman, M.~Kamionkowski and K.~Griest,
  %``Supersymmetric dark matter,''
  Phys.\ Rept.\  {\bf 267}, 195 (1996).
  %%CITATION = PRPLC,267,195;%%

\bibitem{Baltz:1998xv}
  E.~A.~Baltz and J.~Edsjo,
  %``Positron propagation and fluxes from neutralino annihilation in the
  %halo,''
  Phys.\ Rev.\  D {\bf 59}, 023511 (1999).
  %%CITATION = PHRVA,D59,023511;%%


\bibitem{AnnihilationPastWorks}
  E.~A.~Baltz, J.~Edsjo, K.~Freese and P.~Gondolo,
  arXiv:astro-ph/0211239;
  G.~L.~Kane, L.~T.~Wang and T.~T.~Wang,
  Phys.\ Lett.\ B {\bf 536} (2002) 263;
  W.~de Boer, C.~Sander, M.~Horn and D.~Kazakov,
  Nucl.\ Phys.\ Proc.\ Suppl.\  {\bf 113} (2002) 221;
  G.~L.~Kane, L.~T.~Wang and J.~D.~Wells,
  Phys.\ Rev.\ D {\bf 65} (2002) 057701;
  E.~A.~Baltz, J.~Edsjo, K.~Freese and P.~Gondolo,
  Phys.\ Rev.\ D {\bf 65} (2002) 063511;
  D.~Hooper and G.~D.~Kribs,
  Phys.\ Rev.\ D {\bf 70} (2004) 115004;
  S.~Profumo and P.~Ullio,
  JCAP {\bf 0407} (2004) 006.

\bibitem{Hooper:2004bq}
  D.~Hooper and J.~Silk,
  %``Searching for dark matter with future cosmic positron experiments,''
  Phys.\ Rev.\  D {\bf 71}, 083503 (2005).
  %%CITATION = PHRVA,D71,083503;%%

\bibitem{Hisano:2005ec}
  J.~Hisano, S.~Matsumoto, O.~Saito and M.~Senami,
  %``Heavy Wino-like neutralino dark matter annihilation into antiparticles,''
  Phys.\ Rev.\  D {\bf 73}, 055004 (2006);
  %%CITATION = PHRVA,D73,055004;%%
  M.~Asano, S.~Matsumoto, N.~Okada and Y.~Okada,
  %``Cosmic positron signature from dark matter in the littlest Higgs model
  %with T-parity,''
  Phys.\ Rev.\  D {\bf 75}, 063506 (2007)
  %%CITATION = PHRVA,D75,063506;%%

\bibitem{Ando:2005hr}
  S.~Ando,
  Phys.\ Rev.\ Lett.\  {\bf 94}, 171303 (2005).
  %%CITATION = PRLTA,94,171303;%%

\bibitem{Buchmuller:2004nz}
  W.~Buchmuller, P.~Di Bari and M.~Plumacher,
  Annals Phys.\  {\bf 315}, 305 (2005).
  %%CITATION = APNYA,315,305;%%

\bibitem{Giudice:2003jh}
  G.~F.~Giudice, A.~Notari, M.~Raidal, A.~Riotto and A.~Strumia,
  Nucl.\ Phys.\  B {\bf 685}, 89 (2004).
  %%CITATION = NUPHA,B685,89;%%

\bibitem{Kawasaki:2008qe}
  For the recent study, see, for example,
  M.~Kawasaki, K.~Kohri, T.~Moroi and A.~Yotsuyanagi,
  arXiv:0804.3745 [hep-ph].
  %%CITATION = ARXIV:0804.3745;%%

\bibitem{Giudice:1998xp}
  G.~F.~Giudice, M.~A.~Luty, H.~Murayama and R.~Rattazzi,
  JHEP {\bf 9812} (1998) 027.
  %%CITATION = HEP-PH 9810442;%%

\bibitem{Randall:1998uk}
  L.~Randall and R.~Sundrum,
  Nucl.\ Phys.\ B {\bf 557} (1999) 79.
  %%CITATION = HEP-TH 9810155;%%

\bibitem{Moroi:1993mb}
  T.~Moroi, H.~Murayama and M.~Yamaguchi,
  Phys.\ Lett.\  B {\bf 303}, 289 (1993).
  %%CITATION = PHLTA,B303,289;%%

\bibitem{GravFromScalars}
  M.~Endo, K.~Hamaguchi and F.~Takahashi,
  Phys.\ Rev.\ Lett.\  {\bf 96}, 211301 (2006);
  %%CITATION = PRLTA,96,211301;%%
  S.~Nakamura and M.~Yamaguchi,
  Phys.\ Lett.\  B {\bf 638}, 389 (2006);
  %%CITATION = PHLTA,B638,389;%%
  M.~Dine, R.~Kitano, A.~Morisse and Y.~Shirman,
  Phys.\ Rev.\  D {\bf 73}, 123518 (2006);
  %%CITATION = PHRVA,D73,123518;%%
  T.~Asaka, S.~Nakamura and M.~Yamaguchi,
  %``Gravitinos from heavy scalar decay,''
  Phys.\ Rev.\  D {\bf 74}, 023520 (2006);
  %%CITATION = PHRVA,D74,023520;%%
  M.~Endo, K.~Hamaguchi and F.~Takahashi,
  Phys.\ Rev.\  D {\bf 74}, 023531 (2006);
  %%CITATION = PHRVA,D74,023531;%%
  M.~Kawasaki, F.~Takahashi and T.~T.~Yanagida,
  Phys.\ Lett.\  B {\bf 638}, 8 (2006);
  %%CITATION = PHLTA,B638,8;%%
  Phys.\ Rev.\  D {\bf 74}, 043519 (2006);
  %%CITATION = PHRVA,D74,043519;%%
  M.~Endo, F.~Takahashi and T.~T.~Yanagida,
  Phys.\ Rev.\  D {\bf 76}, 083509 (2007);
  %%CITATION = PHRVA,D76,083509;%%
  Phys.\ Lett.\  B {\bf 658}, 236 (2008).
  %%CITATION = PHLTA,B658,236;%%


\bibitem{Roy:1996bua}
  S.~Roy and B.~Mukhopadhyaya,
  %``Some implications of a supersymmetric model with R-parity breaking
  %bilinear interactions,''
  Phys.\ Rev.\  D {\bf 55}, 7020 (1997);
  F.~Takayama and M.~Yamaguchi,
  %``Pattern of neutrino oscillations in supersymmetry with bilinear  R-parity
  %violation,''
  Phys.\ Lett.\  B {\bf 476}, 116 (2000);
  M.~Hirsch, M.~A.~Diaz, W.~Porod, J.~C.~Romao and J.~W.~F.~Valle,
  %``Neutrino masses and mixings from supersymmetry with bilinear R-parity
  %violation: A theory for solar and atmospheric neutrino oscillations,''
  Phys.\ Rev.\  D {\bf 62}, 113008 (2000)
  [Erratum-ibid.\  D {\bf 65}, 119901 (2002)];
  A.~Abada, S.~Davidson and M.~Losada,
  %``Neutrino masses and mixings in the MSSM with soft bilinear R(p)
  %violation,''
  Phys.\ Rev.\  D {\bf 65}, 075010 (2002);
  E.~J.~Chun, D.~W.~Jung and J.~D.~Park,
  %``Bi-large neutrino mixing from bilinear R-parity violation with
  %non-universality,''
  Phys.\ Lett.\  B {\bf 557}, 233 (2003).


\bibitem{seesaw}
    T.\ Yanagida,
    in ``Proceedings of the Workshop on Unified Theory and Baryon
    Number of the Universe,''
    eds.\ O.\ Sawada and A.\ Sugamoto (KEK, Tsukuba, 1979) p.95;
    M.\ Gell-Mann, P.\ Ramond and R.\ Slansky,
    in ``Supergravity,''
    eds.\ P.\ van Niewwenhuizen and D.\ Freedman (North Holland,
    1979);
    S.~L.~Glashow,
    in ``Proceedings of the Carg\'ese Summer Institute on Quarks and
    Leptons,'' (Plenum, 1980) p707.

\bibitem{Campbell:1990fa}
  B.~A.~Campbell, S.~Davidson, J.~R.~Ellis and K.~A.~Olive,
  %``Cosmological baryon asymmetry constraints on extensions of the standard
  %model,''
  Phys.\ Lett.\  B {\bf 256}, 484 (1991);
  W.~Fischler, G.~F.~Giudice, R.~G.~Leigh and S.~Paban,
  %``Constraints On The Baryogenesis Scale From Neutrino Masses,''
  Phys.\ Lett.\  B {\bf 258}, 45 (1991);
  H.~K.~Dreiner and G.~G.~Ross,
  %``Sphaleron Erasure Of Primordial Baryogenesis,''
  Nucl.\ Phys.\  B {\bf 410}, 188 (1993).


\bibitem{Takayama:2000uz}
  F.~Takayama and M.~Yamaguchi,
  %``Gravitino dark matter without R-parity,''
  Phys.\ Lett.\  B {\bf 485}, 388 (2000).

\bibitem{Hinshaw:2008kr}
  G.~Hinshaw {\it et al.}  [WMAP Collaboration],
  arXiv:0803.0732 [astro-ph].
  %%CITATION = ARXIV:0803.0732;%%

%\bibitem{wmap}
%    D.~N.~Spergel {\it et al.},
%    Astrophys.\ J.\ Suppl.\  {\bf 148}, 175 (2003).
%    %%CITATION = ASTRO-PH 0302209;%%

\bibitem{Asaka:1998ju}
  T.~Asaka, J.~Hashiba, M.~Kawasaki and T.~Yanagida,
  %``Spectrum of background X-rays from moduli dark matter,''
  Phys.\ Rev.\  D {\bf 58}, 023507 (1998);
  %%CITATION = PHRVA,D58,023507;%%
  G.~Bertone, W.~Buchmuller, L.~Covi and A.~Ibarra,
  %``Gamma-Rays from Decaying Dark Matter,''
  JCAP {\bf 0711}, 003 (2007).
  %%CITATION = JCAPA,0711,003;%%

\bibitem{NFW}
  J.~F.~Navarro, C.~S.~Frenk and S.~D.~M.~White,
  %``A Universal Density Profile from Hierarchical Clustering,''
  Astrophys.\ J.\  {\bf 490}, 493 (1997).
  %%CITATION = ASJOA,490,493;%%

\bibitem{BoostFactor}
%\bibitem{Silk:1992bh}
  J.~Silk and A.~Stebbins,
  Astrophys.\ J.\  {\bf 411} (1993) 439;
%\bibitem{Bergstrom:1998xh}
  L.~Bergstrom, J.~Edsjo and P.~Gondolo,
  Phys.\ Rev.\ D {\bf 59} (1999) 043506.

\bibitem{Ensslin:1996ep}
  T.~A.~Ensslin, P.~L.~Biermann, P.~P.~Kronberg and X.~P.~Wu,
  %``Cosmic Ray Protons and Magnetic Fields in Clusters of Galaxies and their
  %Cosmological Consequences,''
  Astrophys.\ J.\  {\bf 477}, 560 (1997);
  %%CITATION = ASJOA,477,560;%%

\bibitem{Loeb:2000na}
  A.~Loeb and E.~Waxman,
  %``Gamma-Ray Background from Structure Formation in the Intergalactic
  %Medium,''
  Nature {\bf 405}, 156 (2000);
  %%CITATION = NATUA,405,156;%%
%\bibitem{Miniati:2002hs}
  F.~Miniati,
  %``Inter-galactic Shock Acceleration and the Cosmic Gamma-ray Background,''
  Mon.\ Not.\ Roy.\ Astron.\ Soc.\  {\bf 337}, 199 (2002);
  %%CITATION = MNRAA,337,199;%%

\bibitem{Gao:1990bh}
  Y.~T.~Gao, F.~W.~Stecker, M.~Gleiser and D.~B.~Cline,
  %``LARGE SCALE ANISOTROPY IN THE EXTRAGALACTIC gamma-ray BACKGROUND AS A PROBE
  %FOR COSMOLOGICAL ANTIMATTER,''
  Astrophys.\ J.\  {\bf 361}, L37 (1990);
  %%CITATION = ASJOA,361,L37;%%
%\bibitem{Dolgov:1992pu}
  A.~Dolgov and J.~Silk,
  %``Baryon isocurvature fluctuations at small scales and baryonic dark
  %matter,''
  Phys.\ Rev.\  D {\bf 47}, 4244 (1993).
  %%CITATION = PHRVA,D47,4244;%%

\bibitem{Hawking:1974rv}
  S.~W.~Hawking,
  %``Black hole explosions,''
  Nature {\bf 248}, 30 (1974);
  %%CITATION = NATUA,248,30;%%
  K.~Maki, T.~Mitsui and S.~Orito,
  %``Local Flux of Low-Energy Antiprotons from Evaporating Primordial Black
  %Holes,''
  Phys.\ Rev.\ Lett.\  {\bf 76}, 3474 (1996).
  %%CITATION = PRLTA,76,3474;%%

\bibitem{Stecker:1998ux}
  F.~W.~Stecker,
  %``Intergalactic extinction of high energy gamma-rays,''
  Astropart.\ Phys.\  {\bf 11}, 83 (1999).
  %%CITATION = APHYE,11,83;%%


\bibitem{Moskalenko:1997gh}
  I.~V.~Moskalenko and A.~W.~Strong,
  %``Production and propagation of cosmic-ray positrons and electrons,''
  Astrophys.\ J.\  {\bf 493}, 694 (1998);
  %%CITATION = ASJOA,493,694;%%
  A.~W.~Strong and I.~V.~Moskalenko,
  %``Propagation of cosmic-ray nucleons in the Galaxy,''
  Astrophys.\ J.\  {\bf 509}, 212 (1998);
  %%CITATION = ASJOA,509,212;%%
  A.~W.~Strong, I.~V.~Moskalenko and O.~Reimer,
  %``Diffuse continuum gamma rays from the Galaxy,''
  Astrophys.\ J.\  {\bf 537}, 763 (2000)
  [Erratum-ibid.\  {\bf 541}, 1109 (2000)];
  %%CITATION = ASJOA,537,763;%%

%\cite{Strong:2004ry}
\bibitem{Strong:2004ry}
  A.~W.~Strong, I.~V.~Moskalenko and O.~Reimer,
  %``A new determination of the extragalactic diffuse gamma-ray background from
  %EGRET data,''
  Astrophys.\ J.\  {\bf 613}, 956 (2004).
  %%CITATION = ASJOA,613,956;%%

\bibitem{glast}
  N.~Gehrels and P.~Michelson,
  %``GLAST: The next-generation high energy gamma-ray astronomy mission,''
  Astropart.\ Phys.\  {\bf 11}, 277 (1999).
  %%CITATION = APHYE,11,277;%%

\bibitem{pamela}
  M.~Boezio {\it et al.},
  %``The space experiment PAMELA,''
  Nucl.\ Phys.\ Proc.\ Suppl.\  {\bf 134}, 39 (2004).
  %%CITATION = NUPHZ,134,39;%%

\bibitem{PAMELA08}
  PAMELA Homepage, {\tt http://pamela.roma2.infn.it}.

\bibitem{Ibarra:2008qg}
  A.~Ibarra and D.~Tran,
  %``Antimatter Signatures of Gravitino Dark Matter Decay,''
  arXiv:0804.4596 [astro-ph].
  %%CITATION = ARXIV:0804.4596;%%





\end{thebibliography}
\end{document}